\documentstyle[12pt,amstex,amssymb]{article}
\input{feynman}
\def\Was{W\c as}
\def\Order#1{${\cal O}(#1$)}

\def\alf1{ {\alpha\over\pi} }

\begin{document}
 
\begin{titlepage}
 \begin{flushright}
 
 {\bf UTHEP-95-0801 }\\
 {\bf August 1995}\\
\end{flushright}
\vspace{1cm}
 
\begin{center}
{\LARGE
Gauge Invariant YFS Exponentiation of (Un)stable \\
$W^+W^-$ Production At and Beyond LEP2 Energies$^{\dagger}$
}
\end{center}

\begin{center}
 {\bf S. Jadach}\\
   {\em Institute of Nuclear Physics,
        ul. Kawiory 26a, Krak\'ow, Poland}\\
   {\em CERN, Theory Division, CH-1211 Geneva 23, Switzerland,}\\
 {\bf W.  P\l{a}czek}$^{\star}$\\
   {\em Department of Physics and Astronomy,\\
   The University of Tennessee, Knoxville, Tennessee 37996-1200},\\
 {\bf M. Skrzypek}\\
   {\em Institute of Nuclear Physics,
        ul. Kawiory 26a, Krak\'ow, Poland}\\
 {\bf B.F.L. Ward}\\
   {\em Department of Physics and Astronomy,\\
   The University of Tennessee, Knoxville, Tennessee 37996-1200,\\
   SLAC, Stanford University, Stanford, California 94309}
\end{center}

\vspace{0.75cm}
\begin{center}
{\bf   Abstract}
\end{center}
 
We present the theoretical basis and sample Monte Carlo data for the YFS
exponentiated calculation of $e^+e^- \rightarrow
W^+ W^- \rightarrow  f_1\bar f'_1 + \bar f_2 f'_2$ at and beyond
LEP2 energies, where the left-handed parts of
$f_i$ and $f'_i$ are the respective upper and lower
components of an $SU_{2L}$ doublet, $i=1,2$.
The problem of gauge invariance of the radiation from the unstable
charged spin~1 $W^{\pm}$ is solved in an entirely physical manner. Our
formulas are illustrated in a proto-typical YFS Monte Carlo
event generator YFSWW2, wherein both Standard Model and anomalous
triple gauge boson couplings are allowed.

\vspace{0.5cm}
\begin{center}
{\it To be submitted to Physical Review D}
\end{center}
 
\vspace{1.50cm}
\renewcommand{\baselinestretch}{0.1}
\footnoterule
\noindent
{\footnotesize
\begin{itemize}
\item[${\dagger}$]
Work partly supported by the Polish Government
grants KBN 2P30225206 and 2P03B17210 and
by the US Department of Energy Contracts  DE-FG05-91ER40627
and   DE-AC03-76ER00515.

\item[${}^{\star}$]
On leave of absence from Institute of Computer Science, Jagellonian
University, Krak\'ow, Poland.
\end{itemize}
}
 
\begin{flushleft} 
{\bf UTHEP-95-0801 }\\
{\bf August 1995}\\
\end{flushleft}

\end{titlepage}

\section{Introduction}

The problem of the precision calculation of the
process $e^+e^- \to W^+W^- +n(\gamma)\to 4 fermions+n(\gamma)$
at and beyond LEP2
energies is of considerable interest in connection with the
verification and tests of the $SU_{2L}\times U_1$ model of Glashow,
Salam and Weinberg~\cite{gsw} of the electroweak interaction.
Indeed, these processes are the primary objective of the initial
LEP2 physics program, providing as they do both a window on the most
precise measurement of the W rest mass and a window on the most precise
test of the fundamental non-Abelian triple and quartet gauge field
self-interactions in principle, for example. In this paper, we present
the theoretical basis of
the rigorous Yennie-Frautschi-Suura (YFS)~\cite{yfs}
Monte Carlo approach~\cite{sjw} to
these processes. For completeness, we shall illustrate our results
with a proto-typical Monte Carlo event generator which will be
exact in the infrared regime and will be of leading logarithmic accuracy
through ${\cal O}(\alpha^2)$ in the hard radiative regime. A more
accurate realization of our results will appear elsewhere~\cite{elsewh}.
 
Referring now to the results in Refs.~\cite{dima,frits}, it is clear
that there are problems of principle in carrying through a 
manifestly gauge invariant 
realization of the production and decay of massive
charged spin~1 boson pairs in $e^+e^-$ with the presence of radiation.
Indeed, some controversy did exist in the literature on just how one
should proceed even in the stable particle case~\cite{dima,frits},
where for example the current splitting approach of Ref.~\cite{dima}
has been questioned as to accuracy and appropriateness in
Ref.~\cite{frits}. We note that recently it has been verified~\cite{WWWG}
that the approach in Ref.~\cite{dima} is indeed 
accurate enough for the requirements of the LEP2 physics program
as it is currently envisioned.
We will show that the YFS theory will afford us
an arena in which we can resolve this controversy.
\par
While we were preparing this manuscript, we became aware of independent
and related results by Baur and Zeppenfeld (B-Z)~\cite{B-Z}
for the problem of
quark-pair annihilation into $W^{\pm}$ followed by a lepton-pair
$W^{\pm}$ decay, a process of interest in hadron-hadron collider physics
for example. We will therefore compare our approach with that of B-Z
in what follows. We shall see that the two approaches are consistent
with one another. Moreover, during this same period,
we became aware of an independent derivation in Ref.~\cite{argy}
of the solution to the gauge invariance problem for radiative corrections
to the processes of interest to us here. We will therefore use
the results of Ref.~\cite{argy} in what follows, as they are
in complete agreement with our work insofar as the issue of gauge
invariance is concerned. \par
In addition, in the original YFS paper~\cite{yfs}, there is no explicit
discussion
of charged spin~1 massive radiation. Thus, we will need to extend the
the rigorous YFS theory to this case as well. This means that our
analysis is of theoretical interest in its own right as a study of
the infrared limit of massless vector radiation from massive charged
vector fundamental particles.

Our work is organized as follows. In the next section, we set our
notational conventions. In section~3, we derive the extension of
YFS theory to spin 1 charged particles. In section 4, we derive
the corresponding YFS formula for the processes
$e^+e^- \to W^+W^- +n(\gamma)\to 4 fermions+n(\gamma)$
and show that it is \underline{manifestly gauge invariant}. 
In section~5, we
illustrate our YFS formula with Monte Carlo data based on the
proto-typical MC event generator YFSWW2~\cite{yfsww2}, which uses the Born
level cross section of Ref.~\cite{zep} as an input to achieve
LL accuracy in the hard radiative regime and exactness in the infrared
regime. Section~6 contains our summary remarks and, finally, some useful
formulae are collected in appendices.

\section{Preliminaries}

In this section we review the relevant aspects of our YFS Monte Carlo
methods as they pertain to the problem of extending our methods to
the $W^+W^-$ pair production processes of interest to us here.
In this way we also set our notation and define our kinematics.

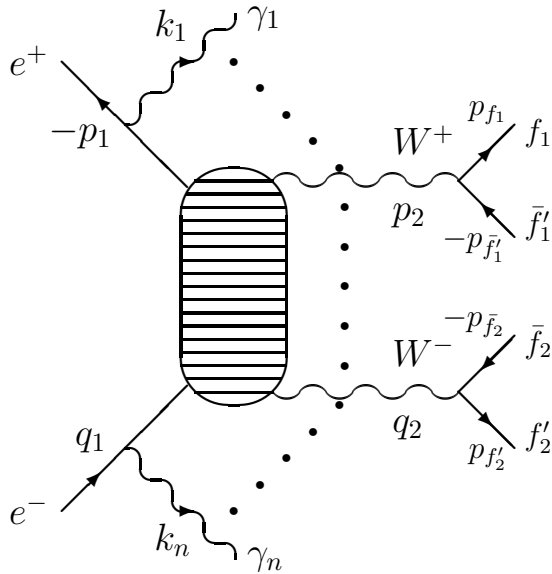
\begin{figure}[!ht]
\centering
\begin{picture}(48000,27000)
\thicklines
\bigphotons
\put(20000,12000){\oval(4000,9000)}
\put(18500,16000){\line(1,0){3000}}
\put(18200,15500){\line(1,0){3600}}
\multiput(18000,15000)(0,-500){13}{\line(1,0){4000}}
\put(18500, 8000){\line(1,0){3000}}
\put(18200, 8500){\line(1,0){3600}}
\THICKLINES
\drawline\fermion[\SE\REG](13500,20500)[6700]
\drawarrow[\NW\ATBASE](15000,19000)
\drawline\photon[\NE\FLIPPEDFLAT](\pmidx,\pmidy)[7]
\drawarrow[\E\ATBASE](18200,20500)
\put(11500,20000){\large $e^+$}
\put(13000,17500){\large $-p_1$}
\put(20500,22000){\large $\gamma_1$}
\put(17000,21500){\large $k_1$}
\drawline\fermion[\NE\REG](13500, 3500)[6700]
\drawarrow[\NE\ATTIP](15000, 5000)
\drawline\photon[\SE\FLAT](\pmidx,\pmidy)[7]
\drawarrow[\E\ATBASE](18200,3500)
\put(11500, 3000){\large $e^-$}
\put(14000, 6000){\large $q_1$}
\put(20500, 1500){\large $\gamma_n$}
\put(17000, 2000){\large $k_n$}
\drawline\photon[\E\REG](21500,16000)[7]
\drawline\fermion[\NE\REG](\photonbackx,\photonbacky)[3000]
\drawarrow[\NE\ATBASE](\pmidx,\pmidy)
\drawline\fermion[\SE\REG](\photonbackx,\photonbacky)[3000]
\drawarrow[\NW\ATBASE](\pmidx,\pmidy)
\put(26000,17000){\large $W^+$}
\put(26000,14500){\large $p_2$}
\put(31000,17500){$f_1$}
\put(31000,14000){$\bar{f_1'}$}
\put(28500,18500){ $p_{f_1}$}
\put(27500,13500){ $-p_{\bar{f_1'}}$}
\drawline\photon[\E\FLIPPED](21500, 8000)[7]
\drawline\fermion[\NE\REG](\photonbackx,\photonbacky)[3000]
\drawarrow[\SW\ATBASE](\pmidx,\pmidy)
\drawline\fermion[\SE\REG](\photonbackx,\photonbacky)[3000]
\drawarrow[\SE\ATBASE](\pmidx,\pmidy)
\put(26000, 9000){\large $W^-$}
\put(26000, 6500){\large $q_2$}
\put(31000, 9500){$\bar{f_2}$}
\put(31000, 6000){$f_2'$}
\put(27500,10500){ $-p_{\bar{f_2}}$}
\put(28500, 5500){ $p_{f_2'}$}
\multiput(20000,20500)(1000,-1000){5}{\circle*{350}}
\multiput(24200, 9000)(   0, 1500){5}{\circle*{350}}
\multiput(20000, 3500)(1000, 1000){5}{\circle*{350}}

\end{picture}

\caption{\small\sf 
The process 
\protect$e^+e^-\rightarrow W^+W^-\rightarrow 4 fermions = f_1+\bar{f'_1}
+\bar{f_2} + f_2'$, where 
\protect$\left({}^{f_i}_{f'_i}\right)$, \protect$i=1,2$, 
are SU$_{2L}$ doublets. Here, $p_A$ is the 4-momentum of $A$, 
\protect$A=f_i,f'_i,$
\protect$p_1(q_1)$ and $p_2(q_2)$ 
are the 4-momenta of \protect$e^+(e^-)$ and \protect$W^+(W^-)$
respectively. We use the notation \protect$C_L\equiv P_L\,C\equiv 
\frac{1}{2}(1-\gamma_5)\,C $ for any $C$.
}
\label{fig:WWprod}
\end{figure}

More precisely, the problem we study herein is illustrated in 
Fig.~\ref{fig:WWprod},
together with the respective kinematics: $e^++e^-\rightarrow W^++W^-
+n(\gamma)\rightarrow 4 fermions + n(\gamma)$ at CMS energies $\sqrt{s}
\ge 2M_W$, so that we may neglect $m_e^2/s$ compared to one.
This corresponds to the case of
LEP2 and of the NLC (level 0 designs of the NLC now are in progress at
several laboratories~\cite{NLC}). Let us focus for the moment on
$W^+W^-$ pair production part of Fig.~\ref{fig:WWprod}.
In Refs.~\cite{yfs2,yfs3}, for the case that the $W$'s are replaced
by the fermion pair $f\bar f$, of rest masses $m_f$
and of charges $\pm Q_fe$,
we have realized by Monte Carlo methods the process $e^+ + e^-\rightarrow
f+\bar{f} + n(\gamma)$ via the fundamental YFS formula
\begin{align}
d\sigma =  e^{2\alpha\Re B + 2\alpha\tilde{B}}
\sum_{n=0}^{\infty} \frac{1}{n!} \int \prod_{j=1}^n 
  \frac{d^3k_j}{k_j^0}\int \frac{d^4y}{(2\pi)^4}
& \, e^{iy(p_1+q_1-p_2-q_2-\sum_j k_j)+D}
\notag\\
& \bar{\beta}_n(k_1,\ldots,k_n)\frac{d^3p_2d^3q_2}{p_2^0q_2^0},
\label{eqone}
\end{align}
where the
real infrared function $\tilde{B}$ and the virtual infrared function $B$ are
given in Refs.~\cite{yfs,yfs2,BHLUMI-89,BHLUMI-92,ward},
and where we note the usual connections
\begin{align}
2\alpha\,\tilde{B} & = \int^{k\le K_{max}}{d^3k\over k_0}\tilde{S}(k),
\notag\\
D & =\int d^3k{\tilde{S}(k)\over k^0}
     \left(e^{-iy k} -\theta(K_{max}-k)\right)
\label{eqtwo}
\end{align} 
for the standard YFS infrared emission factor
\begin{equation}
\tilde S(k)= {\alpha\over 4\pi^2}\left[Q_fQ_{f'}
\left({p_1\over p_1 k}-{q_1 \over q_1 k}\right)^2+\dots\right]
\label{eqthree}
\end{equation} 
if $Q_f$ is the electric charge of $f$ in units of the positron charge. 
Here, the ``$\ldots$'' 
represent the remaining terms in $\tilde S(k)$ obtained from the one
given by respective substitutions of $Q_f$, $p_1$, $Q_{f'}
$, $   q_1$ with corresponding values for the
other pairs of the respective external charged legs according to the YFS
prescription in Ref.~\cite{yfs} (wherein due attention is taken to obtain
the correct relative sign of each of the terms in $\tilde S(k)$ according to
this latter prescription) and in Ref.~\cite{yfs2,yfs3}, $f\ne e$,
$f' = \bar f$.

The YFS hard photon residuals $\bar\beta_i$ in (\ref{eqone}), $i=0,1,2$,
are given in Refs.~\cite{yfs2,yfs3} for YFS2, YFS3 so that these latter
event generators calculate the YFS exponentiated exact
${\cal O}(\alpha^2)$ LL
cross section for $e^+e^-\rightarrow f\bar f+ n(\gamma)$
with multiple initial, (initial+final) state radiation respectively
using a corresponding Monte Carlo realization
of (\ref{eqone}). In the next sections, we use explicit
Feynman diagrammatic methods to extend the realization of (\ref{eqone})
in YFS2, YFS3 to the corresponding Monte Carlo realization of
the respective application of (\ref{eqone}) to
$e^+e^-\rightarrow W^+W^-+n(\gamma)\rightarrow 4 fermions+n(\gamma)$.

\section{YFS theory for massive charged vector particles}

In this section, we present the required formulas for extending
the exact ${\cal O}(\alpha^2)$ LL
Monte Carlo realization of the hard photon residuals $\bar\beta_n$
in YFS2, YFS3 \cite{yfs2,yfs3} to the corresponding Monte Carlo
realization for the $W^+W^-$-pair production processes. We begin by
deriving the respective YFS real and virtual infrared functions
for these latter processes.
\par
Referring to the kinematics in Fig.~\ref{fig:WWprod} and to the definition 
of the YFS infrared functions $\tilde S(k),S(k)$ in Ref.~\cite{yfs},
we see that to extend the infrared YFS calculus to $W$'s it is
enough to show that in the respective infrared regime for an
emitted photon of 4-momentum $k$
the amplitude for spin~1, charge $Q_We$ and mass $M_W$ for the
emitting particle is related to that for spin~1/2, charge $Q_fe$
and rest mass $m_f$ by the substitutions of the respective
Lorentz group representation factor, the charge, the corresponding
4-vectors and radiationless (Born) amplitude. From this result,
it is immediate that the formulas given in Ref.~\cite{yfs} for
for $\tilde S(k),S(k)$ for spin~1/2 hold also for spin~1 with
the corresponding substitution of charges and massive 4-vectors.
Let us now establish this correspondence of the infrared regimes of
spin~1/2 and spin~1 massive, charged particles.

\begin{figure}[!ht]
\centering
\begin{picture}(48000,27000)
\THICKLINES
\bigphotons
\drawline\fermion[\E\REG]( 3000,18000)[12000]
\drawarrow[\E\ATBASE]( 6000,18000)
\drawarrow[\E\ATBASE](12000,18000)
\put( 2000,18500){\large $e^-$}
\put( 5500,16500){\large $p$}
\put(11500,16500){\large $p'$}
\drawline\photon[\N\CURLY](\pmidx,\pmidy)[5]
\drawarrow[\LDIR\ATBASE](8800,\pmidy)
\put( 9500,\pmidy){\large $k$}
\put(\photonbackx,\photonbacky){ $\mu$}
\put( 7500,\photonbacky){\large $\gamma$}
\multiput(16000,18000)(1000,0){3}{\circle*{300}}
\put( 8000,14500){\large\bf (a) }
\drawline\photon[\E\REG](30000,18000)[12]
\drawarrow[\SE\ATBASE](33000,18000)
\drawarrow[\SE\ATBASE](39000,18000)
\put(28000,18500){\large $W^-$}
\put(32500,16500){\large $p$}
\put(38500,16500){\large $p'$}
\drawline\photon[\N\CURLY](\pmidx,\pmidy)[5]
\drawarrow[\LDIR\ATBASE](35800,\pmidy)
\put(36500,\pmidy){\large $k$}
\put(\photonbackx,\photonbacky){ $\mu$}
\put(34500,\photonbacky){\large $\gamma$}
\multiput(43000,18000)(1000,0){3}{\circle*{300}}
\put(35000,14500){\large\bf (a$'$) }
\drawline\fermion[\E\REG]( 3000, 5000)[12000]
\drawarrow[\W\ATBASE]( 6000, 5000)
\drawarrow[\W\ATBASE](12000, 5000)
\put( 2000, 5500){\large $e^+$}
\put( 4500, 3500){\large $-p$}
\put(10500, 3500){\large $-p'$}
\drawline\photon[\N\CURLY](\pmidx,\pmidy)[5]
\drawarrow[\LDIR\ATBASE](8800,\pmidy)
\put( 9500,\pmidy){\large $k$}
\put(\photonbackx,\photonbacky){ $\mu$}
\put( 7500,\photonbacky){\large $\gamma$}
\multiput(16000, 5000)(1000,0){3}{\circle*{300}}
\put( 8000, 1500){\large\bf (b) }
\drawline\photon[\E\REG](30000, 5000)[12]
\drawarrow[\SE\ATBASE](33000, 5000)
\drawarrow[\SE\ATBASE](39000, 5000)
\put(28000, 5500){\large $W^+$}
\put(32500, 3500){\large $p$}
\put(38500, 3500){\large $p'$}
\drawline\photon[\N\CURLY](\pmidx,\pmidy)[5]
\drawarrow[\LDIR\ATBASE](35800,\pmidy)
\put(36500,\pmidy){\large $k$}
\put(\photonbackx,\photonbacky){ $\mu$}
\put(34500,\photonbacky){\large $\gamma$}
\multiput(43000, 5000)(1000,0){3}{\circle*{300}}
\put(35000, 1500){\large\bf (b$'$) }

\end{picture}

\caption{\small\sf 
Infrared emission of a photon of 4-momentum $k$ by 
{\bf (a)} an incoming fermion of mass $m_f$, charge \protect$Q_fe$ and 
4-momentum $p$,
{\bf (a$'$)} an incoming spin~1 charged boson of mass $M_W$, charge 
\protect$Q_We$ and  4-momentum $p$, 
{\bf (b)} an incoming anti-fermion of rest mass $m_f$, charge \protect$-Q_fe$
 and 4-momentum $p$, and 
{\bf (b$'$)} an incoming spin~1 charged boson of rest mass $M_W$, 
charge \protect$-Q_We$ and  4-momentum $p$. Here, the components of $p$ 
of course must change in passing from the fermion cases to the vector boson 
cases if the incoming lines are on-shell, for example.
}
\label{fig:photrad}
\end{figure}
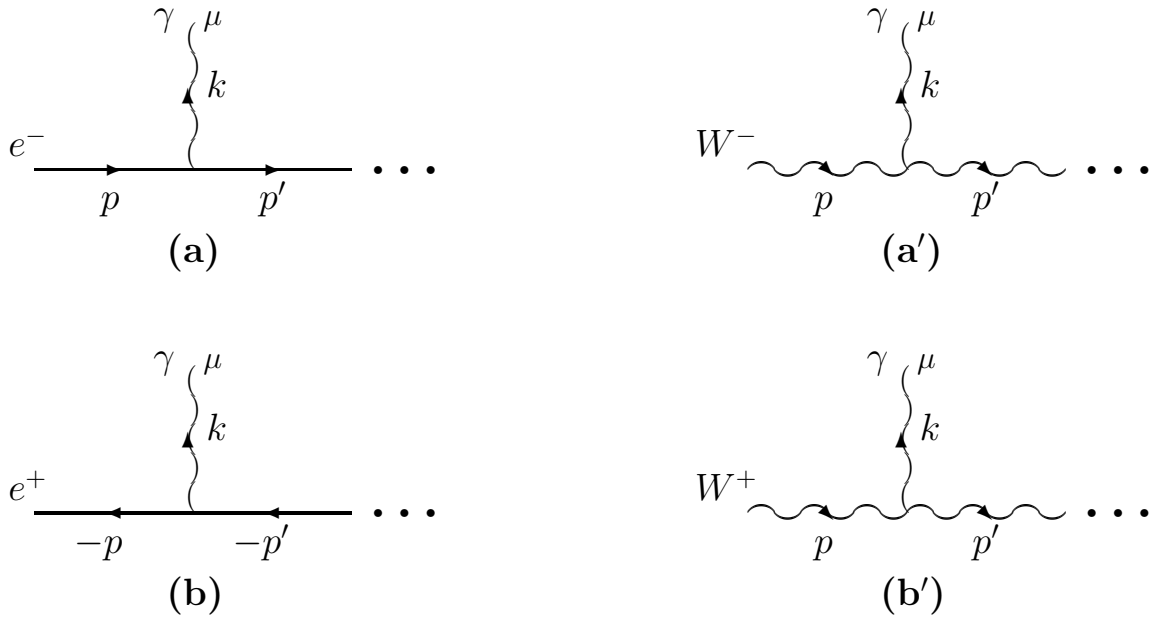

The relevant situations are illustrated in Fig.~\ref{fig:photrad}, where in
{\bf (a)} we have the emission of a photon of 4-momentum $k$ by an incoming
fermion of charge $Q_fe$, rest mass $m_f$, and 4-momentum $p$
to be compared with the analogous situation in
{\bf (a$'$)} where we have the emission of a photon of 4-momentum $k$ by
an incoming spin~1 vector boson $W^-$ of charge $Q_We$, rest mass $M_W$ and
4-momentum $p$. In {\bf (b)}, we have the emission of a photon of 4-momentum
$k$ by an incoming spin~1/2 anti-fermion of charge $-Q_fe$, rest mass
$m_f$, and 4-momentum $p$ to be compared with {\bf (b$'$)} wherein we have the
emission of a photon of 4-momentum $k$ by an incoming spin~1 vector boson
$W^+$ of charge $-Q_We$, rest mass $M_W$ and 4-momentum $p$ as well.
From the standard Feynman methods, for Fig.~\ref{fig:photrad}{\bf (a)}, 
we have the amplitude, for $f=e$ for definiteness, 
\begin{align}
{\cal M}_{2a}&= \cdots {i\over \not\!p'-m_e+i\epsilon}(-iQ_ee\gamma^\mu)u(p)
\notag\\
           &= \cdots {(Q_ee)\over -2kp+k^2+i\epsilon}(\not\!p-\not\!k+m_e)\gamma^\mu u(p)
\notag\\
           &= \cdots {(Q_ee)\over k^2-2kp+i\epsilon}[2p^\mu-k^\mu+i\sigma^{\mu\alpha}k_{\alpha}]u(p) \notag\\
\Rightarrow \lim_{IR} {\cal M}_{2a}&=\cdots {(Q_ee)\over k^2-2kp+i\epsilon}(2p^\mu-k^\mu)u(p),
\label{eq4}
\end{align}
where we define \[\lim_{IR} {\cal A}(k)\equiv\lim_{k\rightarrow 0} {\cal A}(k),
\notag\] for any function ${\cal A}(k)$.
For Fig.~\ref{fig:photrad}{\bf (a$'$)}, we get the corresponding result
\begin{align}
{\cal M}_{2a'} &= \cdots {(-i)(g^{\alpha''\alpha'}-{p'}^{\alpha''}{p'}^{\alpha'}/M_W^2)\over ({p'}^2-M_W^2+i\epsilon)}(iQ_We)
\notag \\
              & [g_{\alpha'\beta}(2p-k)^\mu+g_{\alpha'}^\mu(-p+2k)_\beta+
                g^\mu_\beta(-p'-2k)_{\alpha'}]\epsilon_-^\beta(p) \notag \\
              &=\,\cdots {(Q_We)(g^{\alpha''\alpha'}-{p'}^{\alpha''}{p'}^{\alpha'}/M_W^2)\over (k^2-2kp+i\epsilon)}[g_{\alpha'\beta}(2p-k)^\mu \notag \\
              &\,+2k_\beta g^\mu_{\alpha'}-2k_{\alpha'}g^\mu_\beta-p'_{\alpha'}g^\mu_\beta]\epsilon_-^\beta(p) 
\notag \\
\Rightarrow \lim_{IR} {\cal M}_{2a'}&=\cdots {(Q_We)\over (k^2-2kp+i\epsilon)}(2p^\mu-k^\mu)\epsilon_-^{\alpha''}(p).
\label{eq5}
\end{align}  
Similarly, for Fig.~\ref{fig:photrad}{\bf (b)}, we get the infrared limit as
\begin{align}
{\cal M}_{2b} &=\,
\bar v(p)(-iQ_ee)\gamma^\mu{i\over -\not\!p+\not\!k-m_e+i\epsilon}\cdots\notag\\
              &=\, {\bar v(p)Q_ee\gamma^\mu(-\not\!p+\not\!k+m_e)\over k^2-2kp+i\epsilon}\cdots
\notag\\
              &=\,{Q_ee\bar v(p)[-\gamma^\mu\not\!p-\not\!p\gamma^\mu+(\not\!p+m_e)\gamma^\mu+{1\over 2}(\not\!k\gamma^\mu+\gamma^\mu\not\!k)+{1\over 2}(\gamma^\mu\not\!k-\not\!k\gamma^\mu)]\over k^2-2kp+i\epsilon }\cdots \notag \\
              &=\,{Q_ee\bar v(p)[-2p^\mu+k^\mu-i\sigma^{\mu\alpha}k_\alpha]\over k^2-2kp+i\epsilon }\cdots \notag \\
\Rightarrow \lim_{IR} {\cal M}_{2b}&= {\bar v(p)Q_ee(-2p^\mu+k^\mu)\over k^2-2kp+i\epsilon}\cdots ,
\label{eq6}
\end{align}
whereas for Fig.~\ref{fig:photrad}{\bf (b$'$)} we compute the infrared limit 
as 
\begin{align}
{\cal M}_{2b'} &= \cdots {(-i)(g^{\alpha''\beta}-{p'}^{\alpha''}{p'}^{\beta}/M_W^2)\over ({p'}^2-M_W^2+i\epsilon)}(iQ_We)[\notag \\
              & g_{\beta\alpha}(-2p+k)^\mu+g_{\alpha}^\mu(p'+2k)_\beta+
                g^\mu_\beta(p-2k)_{\alpha}]\epsilon_+^\alpha(p) \notag \\
              &=\,\cdots {(Q_We)(g^{\alpha''\beta}-{p'}^{\alpha''}{p'}^{\beta}/M_W^2)\over (k^2-2kp+i\epsilon)}[g_{\beta\alpha}(-2p+k)^\mu \notag \\
              &\,+2k_\beta g^\mu_{\alpha}-2k_{\alpha}g^\mu_\beta+p'_{\beta}g^\mu_\alpha]\epsilon_+^{\alpha}(p) \notag \\
\Rightarrow \lim_{IR} {\cal M}_{2b'}&=\cdots {(Q_We)\over (k^2-2kp+i\epsilon)}(-2p^\mu+k^\mu)\epsilon_+^{\alpha''}(p).
\label{eq7}
\end{align}  
This shows that the stated correspondence holds.

It follows that we obtain the YFS infrared functions \cite{yfs} $\tilde S(k)$
and $S(k)$ for real and virtual soft photon emission from $W^{+,-}$-lines by 
substituting the respective mass $M_W$ into the corresponding expressions
for these functions for emission from $e^{+,-}$-lines: 
\begin{align}
{\overset{{\tiny(}{\tiny\sim}{\tiny)}}{S}(k)}_{e\bar{e}}|_{m=M_W} & =
{\overset{{\tiny(}{\tiny\sim}{\tiny)}}{S}(k)}_{W^-W^+},  
\notag \\
{\overset{{\tiny(}{\tiny\sim}{\tiny)}}{S}(k)}_{e e}|_{m=M_W} & =
{\overset{{\tiny(}{\tiny\sim}{\tiny)}}{S}(k)}_{W^-W^-}, 
\label{eq8} \\
{\overset{{\tiny(}{\tiny\sim}{\tiny)}}{S}(k)}_{\bar{e}\bar{e}}|_{m=M_W} & =
{\overset{{\tiny(}{\tiny\sim}{\tiny)}}{S}(k)}_{W^+W^+},
\notag 
\end{align}
where the subscripts indicate the respective YFS
infrared functions for $e\bar{e}$, $ee$, $\bar{e}\bar{e}$, $W^-W^+$, $W^-W^-$
and $W^+W^+$ in the obvious manner.

One important point needs to be discussed before we turn to the application
of (\ref{eq8}) to $W^{\pm}$ pair production. This concerns the so-called
Coulomb effect \cite{Coul} --- the enhanced $1/\beta$ behavior of the
${\cal O}(\alpha)$ virtual correction to the Born cross section due
to the exchange of $k\rightarrow 0$ virtual photons, where $\beta$
is the CMS velocity of one of the $W$'s.
Since the YFS
virtual infrared function 
$2\alpha B_{W^-W^+} \equiv \int(d^4k/(k^2-m_\gamma^2+i\epsilon))S(k)_{W^-W^+}$, where $m_\gamma$ is our photon infrared 
regulator mass, 
describes precisely this regime of the virtual photon phase space as well,
we need to remove this Coulomb effect from $B_{W^-W^+}$ so that it can be
treated via the methods of Ref.~\cite{Coul} as accurately as one desires
without double counting it. This we do by defining the analytic subtraction
\begin{align}
B'_{W^-W^+}(\beta)& = B_{W^-W^+}(\beta)-{\theta(\beta_t-\beta)\over\beta}
\lim_{\beta\rightarrow 0}\beta B_{W^-W^+}(\beta)
\notag\\
                  & = B_{W^-W^+}(\beta)-{\pi\over4\beta}\theta(\beta_t-\beta), 
\label{eq88}
\end{align}
where $\theta$ is the usual step function and here we determine
$\beta_t$, the transition velocity which separates the non-relativistic 
regime from the relativistic one insofar as the Coulomb corrections
are concerned,  by the requirement that the ${\cal O}(\alpha^2)$
Coulomb correction, $({\pi\alpha\over\beta})^2/12$, is less than $0.03\%$
for $\beta\ge\beta_t$ --- this gives $\beta_t \cong 0.382$. 
For $\beta\ge\beta_t$,
the Coulomb correction series is a well-behaved part of the usual
perturbation series and does not need special treatment in our analysis.
Our requirement that the ${\cal O}(\alpha^2)$ Coulomb correction
stay below $0.03\%$ for $\beta\ge\beta_t$ is determined with an eye
toward an ultimate goal of $0.1\%$ total precision on our calculations ---
this goal is however beyond the scope of the current paper.
It is apparent that $B'_{W^-W^+}$ and $B_{W^-W^+}$ contain the same
infrared divergences so that we may use $B'_{W^-W^+}$ in our YFS 
exponentiation
algebra without any change in the cancellation of infrared singularities
to all orders in $\alpha$ so that such a use of $B'_{W^-W^+}$ is fully
justified by the original YFS arguments~\cite{yfs} --- the resulting 
cross sections are fully independent of the substitution of $B'_{W^-W^+}$
for $B_{W^-W^+}$ when the theory is summed to all orders in $\alpha$ as
it is proved in the original YFS paper. Thus, we will make this substitution
here and treat the Coulomb effect entirely according to the methods in
Ref.~\cite{Coul}. It can therefore be seen that our procedure for 
arriving at a smoothe transition, in our complete cross section, within
the limits of its physical precision tag, between the Coulomb dominated
$\beta\rightarrow 0$ regime and the relatavistic $\beta\rightarrow 1$ regime
of the $W^{+,-}$ charge form factor is entirely consistent with 
the smoothe interpolation of Schwinger in Ref.~\cite{schw} between
the analogous regimes of the respective charge form factors for
spin 0 and spin ${1\over 2}$ massive charged particles. 

We now turn to the application of the results in this section to the
realization of our YFS Monte Carlo approach to multiple photon radiative
effects in $e^+e^-\rightarrow W^+W^- + n(\gamma)$ at LEP2 and NLC
energies.

\section{Gauge invariant YFS Monte Carlo for 
         $e^+e^-\rightarrow W^+W^- +n(\gamma)$}

In this section we apply the results of the preceding section to develop
a Monte Carlo event generator, YFSWW2~\cite{yfsww2}, which realizes the YFS exponentiated
multiple photon radiation in the process $e^+e^-\rightarrow W^+W^-+n(\gamma)$,
where we will also allow the $W$'s to decay to four-fermion final states.
In the development presented here, we shall work to the leading log 
$\bar\beta_0$
level in the respective YFS hard photon residuals $\bar\beta_n$ in
(\ref{eqone}) as it is applied to $W$-pair production via the substitutions
in (\ref{eq8}).

Specifically, on using the results in (\ref{eq8}), we arrive at the
representation, for the process $e^+e^-\rightarrow W^+W^-+n(\gamma)
\rightarrow f_1+\bar f'_1+f'_2+\bar f_2+n(\gamma)$,
of the fundamental YFS cross section formula
\begin{equation}
d\sigma=e^{2\alpha\,Re\,B+2\alpha\,
\tilde B}\sum_{n=0}^\infty{1\over n!}\int\prod_{j=1}^n{d^3k_j\over k_j^0
}\int{d^4y\over(2\pi)^4}\,e^{iy(p_1+q_1-p_2-q_2-\sum_jk_j)+D}\nonumber\cr
\qquad\bar\beta_n(k_1,\dots,k_n){d^3p_{f_1}d^3p_{\bar{f'}_1}d^3p_{f'_2}d^3p_{\bar{f}_2}
\over p_{f_1}^0p_{\bar f'_1}^0p_{f'_2}^0p_{\bar{f}_2}^0},
\label{eq9}
\end{equation}
where, referring to the kinematics in Fig.~\ref{fig:WWprod}, we have the
identifications  $p_2 = p_{f_1}+p_{\bar f'_1}$, $q_2 = p_{f'_2}+p_{\bar f_2}$,
for the $W^+$, $W^-$ 4-momenta, respectively. The YFS infrared functions
$\tilde B(K_{max})$, $D$, and $B$, by (\ref{eq8}), are obtained from the
the corresponding ones for the process 
$e^+e^-\rightarrow f+\bar f + n(\gamma)$
as given in (\ref{eqthree}) and in Refs.~\cite{yfs2,ward} via the
substitutions: $(Q_f,p_f,m_f)\rightarrow (Q_{W^-},q_2,M_W)$,
$(Q_{\bar f},p_{\bar f},m_f)\rightarrow (Q_{W^+},p_2,M_W)$.
Their analytical representations are given in Appendix~A.
The hard photon residuals $\bar\beta_n$ now contain both the production
and the decay of the $W$'s, which may of course occur either on or off
the $W$ mass-shell. We will work to the $\bar\beta_0$ level and 
we have the identification
\begin{equation}   
\frac{1}{2}\bar\beta_0= \frac{d\sigma_{Born}}{d\Omega_+d\Omega_-},
\label{eq10}
\end{equation}
where $d\Omega_{+(-)}$ is the differential decay solid angle of $f_1(f'_2)$
in the $W^{+(-)}$ rest frame for example. Here, we take the respective
Born cross section, $d\sigma_{Born}$, from Ref.~\cite{zep} for 
definiteness. The result (\ref{eq9}) has been realized via 
the YFS MC methods of two of us, see for example Ref.~\cite{sjw,yfs2},
in the program YFSWW2~\cite{yfsww2}.

\begin{figure}[!ht]
\centering
\begin{picture}(48000,25000)
\bigphotons
\THICKLINES
\drawline\fermion[\E\REG]( 6000,17500)[9000]
\drawarrow[\W\ATBASE](\pmidx,\pmidy)
\put( 4000,17000){\large $e^+$}
\put( 9000,16000){\large $-p_1$}
\drawline\fermion[\S\REG](\pbackx,\pbacky)[12000]
\drawarrow[\N\ATBASE](\pmidx,\pmidy)
\put(13000,\pmidy){\large $\bar{\nu}_e$}
\drawline\fermion[\E\REG]( 6000, 5500)[9000]
\drawarrow[\E\ATBASE](\pmidx,\pmidy)
\put( 4000, 5000){\large $e^-$}
\put(10000, 4000){\large $q_1$}
\drawline\photon[\E\REG](15000,17500)[6]
\put(17500,18500){\large $W^+$}
\put(16500,16000){\large $p_2+k$}
\drawline\fermion[\NE\REG](\photonbackx,\photonbacky)[4000]
\drawarrow[\SW\ATBASE](\pmidx,\pmidy)
\put(21000,19500){$\bar{f_j'}$}
\multiput(22800,\pbacky)(0,-300){20}{\circle*{160}}
\drawline\fermion[\SE\REG](\photonbackx,\photonbacky)[4000]
\drawarrow[\SE\ATBASE](\pmidx,\pmidy)
\put(21000,15000){$f_j$}
\drawline\fermion[\N\REG](\pbackx,\pbacky)[5657]
\drawarrow[\N\ATBASE](\pmidx,\pmidy)
\drawline\photon[\E\REG](\pbackx,\pbacky)[8]
\drawarrow[\E\ATBASE](28300,20600)
\put(\pmidx,18700){\large $k$}
\put(\pmidx,21500){\large $\gamma$}
\drawline\photon[\E\REG](\fermionfrontx,\fermionfronty)[6]
\put(\pmidx,15500){\large $W^+$}
\put(\pmidx,13200){\large $p_2$}
\drawline\fermion[\NE\REG](\photonbackx,\photonbacky)[3000]
\drawarrow[\NE\ATBASE](\pmidx,\pmidy)
\put(32300,16500){$f_1$}
\put(\pfrontx,16800){$p_{f_1}$}
\drawline\fermion[\SE\REG](\photonbackx,\photonbacky)[3000]
\drawarrow[\NW\ATBASE](\pmidx,\pmidy)
\put(32300,12000){$\bar{f_1'}$}
\put(29000,12300){$-p_{\bar{f_1'}}$}
\drawline\photon[\E\FLIPPED](15000, 5500)[6]
\put(17500, 6500){\large $W^-$}
\put(17500, 4000){\large $q_2$}
\drawline\fermion[\NE\REG](\photonbackx,\photonbacky)[3000]
\drawarrow[\SW\ATBASE](\pmidx,\pmidy)
\put(23500, 7500){$\bar{f_2}$}
\put(20200, 8000){$-p_{\bar{f_2}}$}
\drawline\fermion[\SE\REG](\photonbackx,\photonbacky)[3000]
\drawarrow[\SE\ATBASE](\pmidx,\pmidy)
\put(23500,3000){$f_2'$}
\put(\pfrontx,3200){$p_{f_2'}$}
\put(37000,10900){\Huge\bf +}
\multiput(41000,11500)(1000,0){3}{\circle*{350}}

\end{picture}

\caption{\small\sf 
The imaginary parts necessary to include in the amplitude for
\protect$e^+e^-\rightarrow W^+W^-\rightarrow 4 fermions=f_1+\bar{f'_1}
+\bar{f_2} + f_2'$ in order to maintain gauge invariance in the 
presence of radiation by the $W$'s themselves. Here, the notation
is identical to that in Fig.~\ref{fig:WWprod} and the vertical dotted line 
indicates the standard Bjorken-Landau-Cutkosky isolation of the respective
imaginary part.
}
\label{fig:WWfloop}
\end{figure}
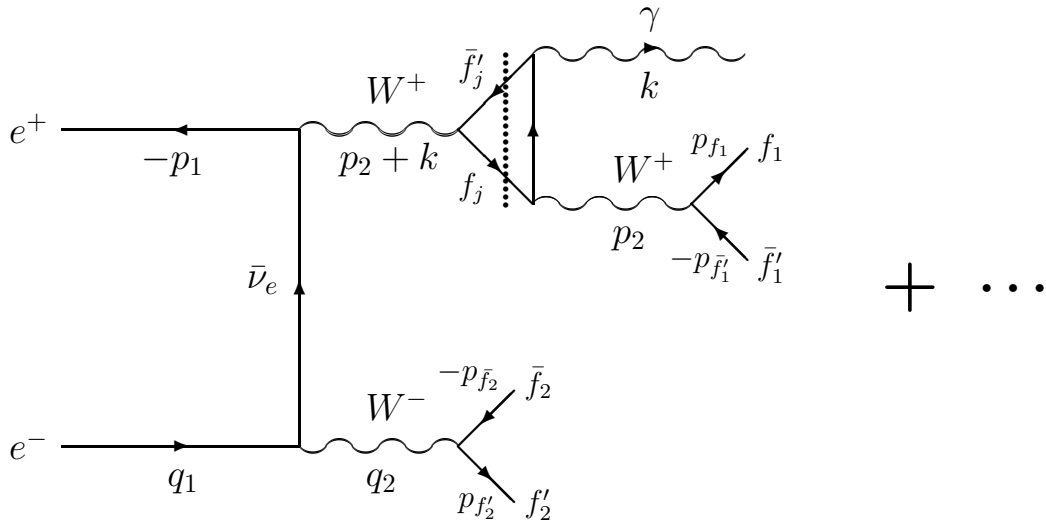

Before we illustrate the type of predictions which we make with YFSWW2
for the LEP2/NLC physics scenarios, let us address one further
important theoretical point concerning the fundamental result (\ref{eq9}).
Specifically, the reader may note that the $W^{\pm}$ decay width $\Gamma_W$
does not appear in the YFS infrared functions $B$ and $\tilde B$ as they
are given via the result (\ref{eq8}). Yet, evidently, when a $W^{\pm}$
of 4-momentum $p'=p+k$ emits radiation of 4-momentum $k$,
the propagator denominator that would most naively
be present immediately preceding the emission vertex would be
$D_W(p') = {p'}^2-M_W^2+ i {p'}^2\Gamma_W/M_W$ and for $p^2=M_W^2$
this does not reduce to the YFS \cite{yfs} infrared algebraic form $k^2+2kp$.
The attendant problems with electromagnetic gauge invariance are now
well-known \cite{dima,frits,B-Z,argy}. The solution to this
apparent dilemma is already presented in Refs.~\cite{B-Z,argy}.
Whenever the $W^{\pm}$ radiate, one can include their decay width
in their propagation provided one also allows the radiated photon
to couple to charges in the graphs which generate the non-zero width
to its respective order as well, see Refs.~\cite{B-Z,argy} for
a detailed discussion of these graphs, which are illustrated in 
Fig.~\ref{fig:WWfloop} 
for the case that $\Gamma_W$ is computed to \Order{\alpha} for example.
The net effect, as explained in detail in Refs.~\cite{B-Z,argy}
and as we have independently verified, is that the
emission vertex for the photon in question is multiplied by
a factor which replaces the pre-emission denominator
$D_W(p') = {p'}^2-M_W^2+ i {p'}^2\Gamma_W/M_W$ for $p^2=M_W^2$
with the desired YFS infrared algebraic form $k^2+2kp$, thereby
maintaining electromagnetic gauge invariance.

We turn now to sample Monte Carlo data from
YFSWW2 in the LEP2 and NLC type energy regimes. This we do in the
next section.

\section{Sample Monte Carlo data for $e^+e^-\rightarrow W^+W^-+n(\gamma)
\rightarrow 4fermions + n(\gamma)$}

In this section, we present sample Monte Carlo data for the process
$e^+e^-\rightarrow W^+W^-+n(\gamma)\rightarrow 4 fermions + n(\gamma)$
in the LEP2/NLC type energy regimes. We have used our Monte Carlo
event generator YFSWW2 as presented above and we have had in mind
in particular
the illustration, among other things,
of the effect of the $W^{\pm}$ contribution
to the YFS real and virtual infrared functions in (\ref{eq9}),
as this effect has not been presented elsewhere.

\begin{table}[!ht]

\centering
\begin{tabular}{||c|c|c|c||}
\hline\hline
 $E_{CM}\,[GeV]$  &  ISR & ISR $+$ Coul.~corr. & 
 ISR $+$ Coul.~corr. $+\,Y'$-corr.\\
\hline\hline
$175$ & $0.4906\pm 0.0002$ & $0.5046\pm 0.0002$ & $0.5053\pm 0.0002$ \\
      & $0.4898\pm 0.0002$ & $0.5037\pm 0.0002$ & $0.5048\pm 0.0002$ \\
\hline
$190$ & $0.6060\pm 0.0007$ & $0.6193\pm 0.0007$ & $0.6217\pm 0.0007$ \\
      & $0.6034\pm 0.0007$ & $0.6166\pm 0.0007$ & $0.6195\pm 0.0007$ \\
\hline
$205$ & $0.6359\pm 0.0008$ & $0.6480\pm 0.0008$ & $0.6514\pm 0.0008$ \\
      & $0.6315\pm 0.0008$ & $0.6436\pm 0.0008$ & $0.6476\pm 0.0008$ \\
\hline\hline
$500$ & $0.2910\pm 0.0003$ & $0.2946\pm 0.0003$ & $0.2970\pm 0.0004$ \\
      & $0.3538\pm 0.0004$ & $0.3582\pm 0.0004$ & $0.3591\pm 0.0004$ \\
\hline\hline
\end{tabular}

\caption{\small\sf The results of the $10^5$ (except for $E_{CM}=175\,GeV$,
where it is $10^6$) statistics sample
from YFSWW2 for the total cross section $\sigma\,[pb]$. 
The upper results at each value of energy are for Standard Model
couplings constants, while the lower ones are for anomalous couplings
constants (\protect$\delta\kappa=\delta\lambda=0.1$). 
See the text for more details.
}
\label{tab:xsec}
\end{table}

Specifically, we will use the 4-fermion final state $f_1=c$, $f'_1=s$,
$f_2=\nu_e$, $f'_2=e$ for definiteness. With an eye toward studies of the
triple gauge boson couplings in the standard model $SU_2\times U_1$
theory of Glashow, Salam and Weinberg~\cite{gsw}, we further follow
the notation of Ref.~\cite{zep} and compute our results for
no anomalous 3-gauge boson couplings and for the values
$\delta\kappa=0.1,\,\delta\lambda=0.1$ for the deviations of the
coupling parameters $\kappa,\lambda$ away from their tree-level
standard model expectations of $1.0,\,0.0$, respectively, in the notation of
Ref.~\cite{zep}. This we do for three LEP2 type CMS energies
$\sqrt s = 175,\,190,\,205\,GeV$ and for the NLC-type energy $500\,GeV$.
The cross sections which we obtain from our simulation
with YFSWW2 were each determined 
from $10^5$ events (except for $175\,GeV$, where we had $10^6$ events)
and for the full solid angle acceptance for each of
the 4-fermions in the final state. The results were computed for both
the pure initial state YFS exponentiated $\bar\beta_0$-level case,
as in Ref.~\cite{KORALW}, referred to as the ISR case in the following,
and in the case when the $W^{\pm}$ contribution
to the soft YFS exponentiated radiative effects is treated exactly,
which we refer to as the ISR+Y$'$ case in the following. Furthermore,
for definiteness, and clarity, the effect of the Coulomb correction,
as given in Ref.~\cite{Coul}, is also illustrated in our results,
both for the simple YFS form factor ISR case and for the full YFS
form factor ISR+Y$'$ case --- the presence of the Coulomb correction
is indicated by ``Coul.'' in the following.
The corresponding results are illustrated in Table~\ref{tab:xsec}.

\begin{figure}[!ht]

\centering
\setlength{\unitlength}{0.1mm}
\begin{picture}(1600,1400)
\put(300,150){\begin{picture}( 1200,1200)
\put(0,0){\framebox( 1200,1200){ }}
\multiput(  190.99,0)(  190.99,0){   6}{\line(0,1){25}}
\multiput(     .00,0)(   19.10,0){  63}{\line(0,1){10}}
\multiput(  190.99,1200)(  190.99,0){   6}{\line(0,-1){25}}
\multiput(     .00,1200)(   19.10,0){  63}{\line(0,-1){10}}
\put( 191,-25){\makebox(0,0)[t]{\large $    0.5 $}}
\put( 382,-25){\makebox(0,0)[t]{\large $    1.0 $}}
\put( 573,-25){\makebox(0,0)[t]{\large $    1.5 $}}
\put( 764,-25){\makebox(0,0)[t]{\large $    2.0 $}}
\put( 955,-25){\makebox(0,0)[t]{\large $    2.5 $}}
\put(1146,-25){\makebox(0,0)[t]{\large $    3.0 $}}
\put(1046,-85){\makebox(0,0)[t]{\Large $ \theta\;[rad] $}}
\multiput(0,     .00)(0,  200.00){   7}{\line(1,0){25}}
\multiput(0,   20.00)(0,   20.00){  60}{\line(1,0){10}}
\multiput(1200,     .00)(0,  200.00){   7}{\line(-1,0){25}}
\multiput(1200,   20.00)(0,   20.00){  60}{\line(-1,0){10}}
\put(-25,   0){\makebox(0,0)[r]{\large $    0.0 $}}
\put(-25, 200){\makebox(0,0)[r]{\large $    0.1 $}}
\put(-25, 400){\makebox(0,0)[r]{\large $    0.2 $}}
\put(-25, 600){\makebox(0,0)[r]{\large $    0.3 $}}
\put(-25, 800){\makebox(0,0)[r]{\large $    0.4 $}}
\put(-25,1000){\makebox(0,0)[r]{\large $    0.5 $}}
\put(-25,1150){\makebox(0,0)[r]{\Large $ \frac{d\sigma}{d\theta}\,
                                         [\frac{pb}{rad}] $}}
\put(350,850){\begin{picture}( 300,900)
 \put(-20,250){\line(1,0){40} }
 \put(50,250){\makebox(0,0)[l]{ISR }}
 \put( 0,200){\makebox(0,0){$\times$}}                                 
 \put(50,200){\makebox(0,0)[l]{ISR $+$ Coul.~corr. }} 
 \put( 0,150){\makebox(0,0){$\star$}}                 
 \put(50,150){\makebox(0,0)[l]{ISR $+$ Coul.~corr. $+\,Y'$-corr. }}
 \put( 0,100){\makebox(0,0){$\diamond$}}                    
 \put(50,100){\makebox(0,0)[l]{ISR $+$ Coul.~corr. $+\,Y'$-corr.
                               $+$ An.~c.~c. }}   
\end{picture}} 
\end{picture}}
\put(300,150){\begin{picture}( 1200,1200)
\thinlines 
\newcommand{\x}[3]{\put(#1,#2){\line(1,0){#3}}}
\newcommand{\y}[3]{\put(#1,#2){\line(0,1){#3}}}
\newcommand{\z}[3]{\put(#1,#2){\line(0,-1){#3}}}
\newcommand{\e}[3]{\put(#1,#2){\line(0,1){#3}}}
\y{   0}{   0}{  70}\x{   0}{  70}{  24} \e{  12}{   66}{   8}
\y{  24}{  70}{ 150}\x{  24}{ 220}{  24} \e{  36}{  213}{  14}
\y{  48}{ 220}{ 126}\x{  48}{ 346}{  24} \e{  60}{  337}{  16}
\y{  72}{ 346}{ 123}\x{  72}{ 469}{  24} \e{  84}{  460}{  20}
\y{  96}{ 469}{ 107}\x{  96}{ 576}{  24} \e{ 108}{  566}{  22}
\y{ 120}{ 576}{ 106}\x{ 120}{ 682}{  24} \e{ 132}{  670}{  22}
\y{ 144}{ 682}{  78}\x{ 144}{ 760}{  24} \e{ 156}{  748}{  24}
\y{ 168}{ 760}{  38}\x{ 168}{ 798}{  24} \e{ 180}{  785}{  24}
\y{ 192}{ 798}{  36}\x{ 192}{ 834}{  24} \e{ 204}{  821}{  26}
\z{ 216}{ 834}{   6}\x{ 216}{ 828}{  24} \e{ 228}{  815}{  26}
\y{ 240}{ 828}{  21}\x{ 240}{ 849}{  24} \e{ 252}{  836}{  26}
\z{ 264}{ 849}{  22}\x{ 264}{ 827}{  24} \e{ 276}{  815}{  26}
\z{ 288}{ 827}{   7}\x{ 288}{ 820}{  24} \e{ 300}{  807}{  26}
\z{ 312}{ 820}{  51}\x{ 312}{ 769}{  24} \e{ 324}{  757}{  24}
\z{ 336}{ 769}{   3}\x{ 336}{ 766}{  24} \e{ 348}{  754}{  24}
\z{ 360}{ 766}{  62}\x{ 360}{ 704}{  24} \e{ 372}{  692}{  24}
\z{ 384}{ 704}{   8}\x{ 384}{ 696}{  24} \e{ 396}{  684}{  24}
\z{ 408}{ 696}{  58}\x{ 408}{ 638}{  24} \e{ 420}{  627}{  22}
\z{ 432}{ 638}{  14}\x{ 432}{ 624}{  24} \e{ 444}{  613}{  22}
\z{ 456}{ 624}{  52}\x{ 456}{ 572}{  24} \e{ 468}{  561}{  20}
\z{ 480}{ 572}{  32}\x{ 480}{ 540}{  24} \e{ 492}{  530}{  20}
\z{ 504}{ 540}{  20}\x{ 504}{ 520}{  24} \e{ 516}{  510}{  20}
\z{ 528}{ 520}{  34}\x{ 528}{ 486}{  24} \e{ 540}{  477}{  20}
\z{ 552}{ 486}{  33}\x{ 552}{ 453}{  24} \e{ 564}{  444}{  18}
\z{ 576}{ 453}{  36}\x{ 576}{ 417}{  24} \e{ 588}{  408}{  18}
\z{ 600}{ 417}{  14}\x{ 600}{ 403}{  24} \e{ 612}{  394}{  18}
\z{ 624}{ 403}{  49}\x{ 624}{ 354}{  24} \e{ 636}{  346}{  16}
\z{ 648}{ 354}{   9}\x{ 648}{ 345}{  24} \e{ 660}{  337}{  16}
\z{ 672}{ 345}{  43}\x{ 672}{ 302}{  24} \e{ 684}{  295}{  16}
\z{ 696}{ 302}{  15}\x{ 696}{ 287}{  24} \e{ 708}{  280}{  14}
\z{ 720}{ 287}{  20}\x{ 720}{ 267}{  24} \e{ 732}{  260}{  14}
\z{ 744}{ 267}{  22}\x{ 744}{ 245}{  24} \e{ 756}{  238}{  14}
\z{ 768}{ 245}{  21}\x{ 768}{ 224}{  24} \e{ 780}{  218}{  14}
\z{ 792}{ 224}{  26}\x{ 792}{ 198}{  24} \e{ 804}{  192}{  12}
\z{ 816}{ 198}{  16}\x{ 816}{ 182}{  24} \e{ 828}{  176}{  12}
\z{ 840}{ 182}{   1}\x{ 840}{ 181}{  24} \e{ 852}{  175}{  12}
\z{ 864}{ 181}{  23}\x{ 864}{ 158}{  24} \e{ 876}{  153}{  12}
\z{ 888}{ 158}{  27}\x{ 888}{ 131}{  24} \e{ 900}{  126}{  10}
\z{ 912}{ 131}{   2}\x{ 912}{ 129}{  24} \e{ 924}{  124}{  10}
\z{ 936}{ 129}{  19}\x{ 936}{ 110}{  24} \e{ 948}{  105}{  10}
\z{ 960}{ 110}{   7}\x{ 960}{ 103}{  24} \e{ 972}{   98}{   8}
\z{ 984}{ 103}{  19}\x{ 984}{  84}{  24} \e{ 996}{   80}{   8}
\z{1008}{  84}{   8}\x{1008}{  76}{  24} \e{1020}{   72}{   8}
\z{1032}{  76}{  11}\x{1032}{  65}{  24} \e{1044}{   61}{   8}
\z{1056}{  65}{   8}\x{1056}{  57}{  24} \e{1068}{   53}{   6}
\z{1080}{  57}{  16}\x{1080}{  41}{  24} \e{1092}{   39}{   6}
\z{1104}{  41}{   3}\x{1104}{  38}{  24} \e{1116}{   35}{   6}
\z{1128}{  38}{  16}\x{1128}{  22}{  24} \e{1140}{   20}{   4}
\z{1152}{  22}{   6}\x{1152}{  16}{  24} \e{1164}{   15}{   4}
\z{1176}{  16}{   7}\x{1176}{   9}{  24} \e{1188}{    8}{   2}
\end{picture}} 
\put(300,150){\begin{picture}( 1200,1200)
\newcommand{\R}[2]{\put(#1,#2){\makebox(0,0){$\times$}}}
\newcommand{\e}[3]{\put(#1,#2){\line(0,1){#3}}}
\R{  12}{  72} \e{  12}{   69}{   8} \R{  36}{ 225} \e{  36}{  219}{  14}
\R{  60}{ 356} \e{  60}{  347}{  16} \R{  84}{ 484} \e{  84}{  474}{  20} 
\R{ 108}{ 590} \e{ 108}{  579}{  22} \R{ 132}{ 697} \e{ 132}{  685}{  24}
\R{ 156}{ 776} \e{ 156}{  763}{  24} \R{ 180}{ 811} \e{ 180}{  798}{  26}
\R{ 204}{ 850} \e{ 204}{  837}{  26} \R{ 228}{ 843} \e{ 228}{  830}{  26}
\R{ 252}{ 869} \e{ 252}{  856}{  26} \R{ 276}{ 846} \e{ 276}{  833}{  26}
\R{ 300}{ 837} \e{ 300}{  824}{  26} \R{ 324}{ 786} \e{ 324}{  774}{  24}
\R{ 348}{ 782} \e{ 348}{  770}{  24} \R{ 372}{ 722} \e{ 372}{  710}{  24}
\R{ 396}{ 709} \e{ 396}{  697}{  24} \R{ 420}{ 649} \e{ 420}{  638}{  22} 
\R{ 444}{ 640} \e{ 444}{  629}{  22} \R{ 468}{ 584} \e{ 468}{  573}{  22}
\R{ 492}{ 552} \e{ 492}{  541}{  20} \R{ 516}{ 533} \e{ 516}{  523}{  20}
\R{ 540}{ 492} \e{ 540}{  482}{  20} \R{ 564}{ 464} \e{ 564}{  455}{  20}
\R{ 588}{ 426} \e{ 588}{  417}{  18} \R{ 612}{ 412} \e{ 612}{  403}{  18}
\R{ 636}{ 362} \e{ 636}{  354}{  16} \R{ 660}{ 352} \e{ 660}{  344}{  16}
\R{ 684}{ 309} \e{ 684}{  301}{  16} \R{ 708}{ 292} \e{ 708}{  285}{  16}
\R{ 732}{ 275} \e{ 732}{  268}{  14} \R{ 756}{ 252} \e{ 756}{  244}{  14}
\R{ 780}{ 228} \e{ 780}{  221}{  14} \R{ 804}{ 203} \e{ 804}{  197}{  12}
\R{ 828}{ 186} \e{ 828}{  180}{  12} \R{ 852}{ 184} \e{ 852}{  178}{  12}
\R{ 876}{ 162} \e{ 876}{  157}{  12} \R{ 900}{ 132} \e{ 900}{  127}{  10}
\R{ 924}{ 130} \e{ 924}{  125}{  10} \R{ 948}{ 112} \e{ 948}{  107}{  10}
\R{ 972}{ 105} \e{ 972}{  101}{  10} \R{ 996}{  87} \e{ 996}{   83}{   8}
\R{1020}{  77} \e{1020}{   73}{   8} \R{1044}{  66} \e{1044}{   63}{   8}
\R{1068}{  58} \e{1068}{   55}{   6} \R{1092}{  41} \e{1092}{   38}{   6}
\R{1116}{  39} \e{1116}{   37}{   6} \R{1140}{  21} \e{1140}{   19}{   4}
\R{1164}{  17} \e{1164}{   15}{   4} \R{1188}{   9} \e{1188}{    8}{   2}
\end{picture}} 
\put(300,150){\begin{picture}( 1200,1200)
\newcommand{\R}[2]{\put(#1,#2){\makebox(0,0){$\star$}}}
\newcommand{\e}[3]{\put(#1,#2){\line(0,1){#3}}}
\R{  12}{  77} \e{  12}{   73}{   8} \R{  36}{ 230} \e{  36}{  223}{  14}
\R{  60}{ 370} \e{  60}{  361}{  18} \R{  84}{ 497} \e{  84}{  487}{  20}
\R{ 108}{ 603} \e{ 108}{  592}{  22} \R{ 132}{ 717} \e{ 132}{  705}{  24}
\R{ 156}{ 788} \e{ 156}{  776}{  24} \R{ 180}{ 818} \e{ 180}{  806}{  26}
\R{ 204}{ 865} \e{ 204}{  852}{  26} \R{ 228}{ 860} \e{ 228}{  847}{  26}
\R{ 252}{ 873} \e{ 252}{  860}{  26} \R{ 276}{ 860} \e{ 276}{  847}{  26}
\R{ 300}{ 841} \e{ 300}{  828}{  26} \R{ 324}{ 784} \e{ 324}{  771}{  24}
\R{ 348}{ 786} \e{ 348}{  774}{  24} \R{ 372}{ 726} \e{ 372}{  714}{  24}
\R{ 396}{ 711} \e{ 396}{  699}{  24} \R{ 420}{ 655} \e{ 420}{  644}{  22}
\R{ 444}{ 638} \e{ 444}{  627}{  22} \R{ 468}{ 586} \e{ 468}{  575}{  22}
\R{ 492}{ 546} \e{ 492}{  535}{  20} \R{ 516}{ 532} \e{ 516}{  522}{  20}
\R{ 540}{ 489} \e{ 540}{  479}{  20} \R{ 564}{ 460} \e{ 564}{  451}{  20}
\R{ 588}{ 422} \e{ 588}{  413}{  18} \R{ 612}{ 406} \e{ 612}{  397}{  18}
\R{ 636}{ 357} \e{ 636}{  349}{  16} \R{ 660}{ 349} \e{ 660}{  341}{  16}
\R{ 684}{ 303} \e{ 684}{  295}{  16} \R{ 708}{ 291} \e{ 708}{  284}{  16}
\R{ 732}{ 269} \e{ 732}{  262}{  14} \R{ 756}{ 250} \e{ 756}{  243}{  14}
\R{ 780}{ 224} \e{ 780}{  218}{  14} \R{ 804}{ 199} \e{ 804}{  192}{  12}
\R{ 828}{ 187} \e{ 828}{  181}{  12} \R{ 852}{ 178} \e{ 852}{  172}{  12}
\R{ 876}{ 155} \e{ 876}{  150}{  12} \R{ 900}{ 130} \e{ 900}{  125}{  10}
\R{ 924}{ 128} \e{ 924}{  123}{  10} \R{ 948}{ 110} \e{ 948}{  106}{  10}
\R{ 972}{ 105} \e{ 972}{  100}{  10} \R{ 996}{  87} \e{ 996}{   83}{   8}
\R{1020}{  77} \e{1020}{   73}{   8} \R{1044}{  63} \e{1044}{   60}{   8}
\R{1068}{  57} \e{1068}{   54}{   6} \R{1092}{  40} \e{1092}{   38}{   6}
\R{1116}{  40} \e{1116}{   37}{   6} \R{1140}{  22} \e{1140}{   20}{   4}
\R{1164}{  17} \e{1164}{   15}{   4} \R{1188}{   9} \e{1188}{    7}{   2}
\end{picture}} 
\put(300,150){\begin{picture}( 1200,1200)
\newcommand{\R}[2]{\put(#1,#2){\makebox(0,0){$\diamond$}}}
\newcommand{\e}[3]{\put(#1,#2){\line(0,1){#3}}}
\R{  12}{  82} \e{  12}{   78}{   8} \R{  36}{ 249} \e{  36}{  242}{  14}
\R{  60}{ 391} \e{  60}{  383}{  18} \R{  84}{ 528} \e{  84}{  518}{  20}
\R{ 108}{ 642} \e{ 108}{  631}{  22} \R{ 132}{ 756} \e{ 132}{  744}{  24}
\R{ 156}{ 833} \e{ 156}{  820}{  26} \R{ 180}{ 863} \e{ 180}{  850}{  26}
\R{ 204}{ 904} \e{ 204}{  891}{  26} \R{ 228}{ 907} \e{ 228}{  893}{  26}
\R{ 252}{ 908} \e{ 252}{  895}{  26} \R{ 276}{ 890} \e{ 276}{  877}{  26}
\R{ 300}{ 867} \e{ 300}{  854}{  26} \R{ 324}{ 809} \e{ 324}{  797}{  26}
\R{ 348}{ 800} \e{ 348}{  788}{  26} \R{ 372}{ 740} \e{ 372}{  727}{  24}
\R{ 396}{ 723} \e{ 396}{  711}{  24} \R{ 420}{ 657} \e{ 420}{  645}{  22}
\R{ 444}{ 638} \e{ 444}{  627}{  22} \R{ 468}{ 584} \e{ 468}{  573}{  22} 
\R{ 492}{ 543} \e{ 492}{  533}{  20} \R{ 516}{ 522} \e{ 516}{  512}{  20}
\R{ 540}{ 474} \e{ 540}{  465}{  20} \R{ 564}{ 441} \e{ 564}{  431}{  18}
\R{ 588}{ 405} \e{ 588}{  396}{  18} \R{ 612}{ 389} \e{ 612}{  381}{  18}
\R{ 636}{ 331} \e{ 636}{  323}{  16} \R{ 660}{ 324} \e{ 660}{  316}{  16}
\R{ 684}{ 279} \e{ 684}{  272}{  14} \R{ 708}{ 259} \e{ 708}{  252}{  14} 
\R{ 732}{ 242} \e{ 732}{  235}{  14} \R{ 756}{ 219} \e{ 756}{  213}{  14}
\R{ 780}{ 195} \e{ 780}{  189}{  12} \R{ 804}{ 182} \e{ 804}{  176}{  12} 
\R{ 828}{ 159} \e{ 828}{  153}{  12} \R{ 852}{ 153} \e{ 852}{  147}{  10}
\R{ 876}{ 128} \e{ 876}{  123}{  10} \R{ 900}{ 108} \e{ 900}{  103}{  10}
\R{ 924}{ 104} \e{ 924}{  100}{  10} \R{ 948}{  90} \e{ 948}{   85}{   8}
\R{ 972}{  84} \e{ 972}{   80}{   8} \R{ 996}{  69} \e{ 996}{   65}{   8}
\R{1020}{  59} \e{1020}{   56}{   6} \R{1044}{  49} \e{1044}{   46}{   6}
\R{1068}{  45} \e{1068}{   42}{   6} \R{1092}{  29} \e{1092}{   27}{   4}
\R{1116}{  28} \e{1116}{   26}{   4} \R{1140}{  17} \e{1140}{   15}{   4}
\R{1164}{  13} \e{1164}{   12}{   4} \R{1188}{   8} \e{1188}{    7}{   2}
\end{picture}} 

\end{picture} 

\caption{\small\sf $W^-$ angular distributions for $E_{CM}=190\,GeV$. 
See the text for more details.
}
\label{fig:e190}
\end{figure}
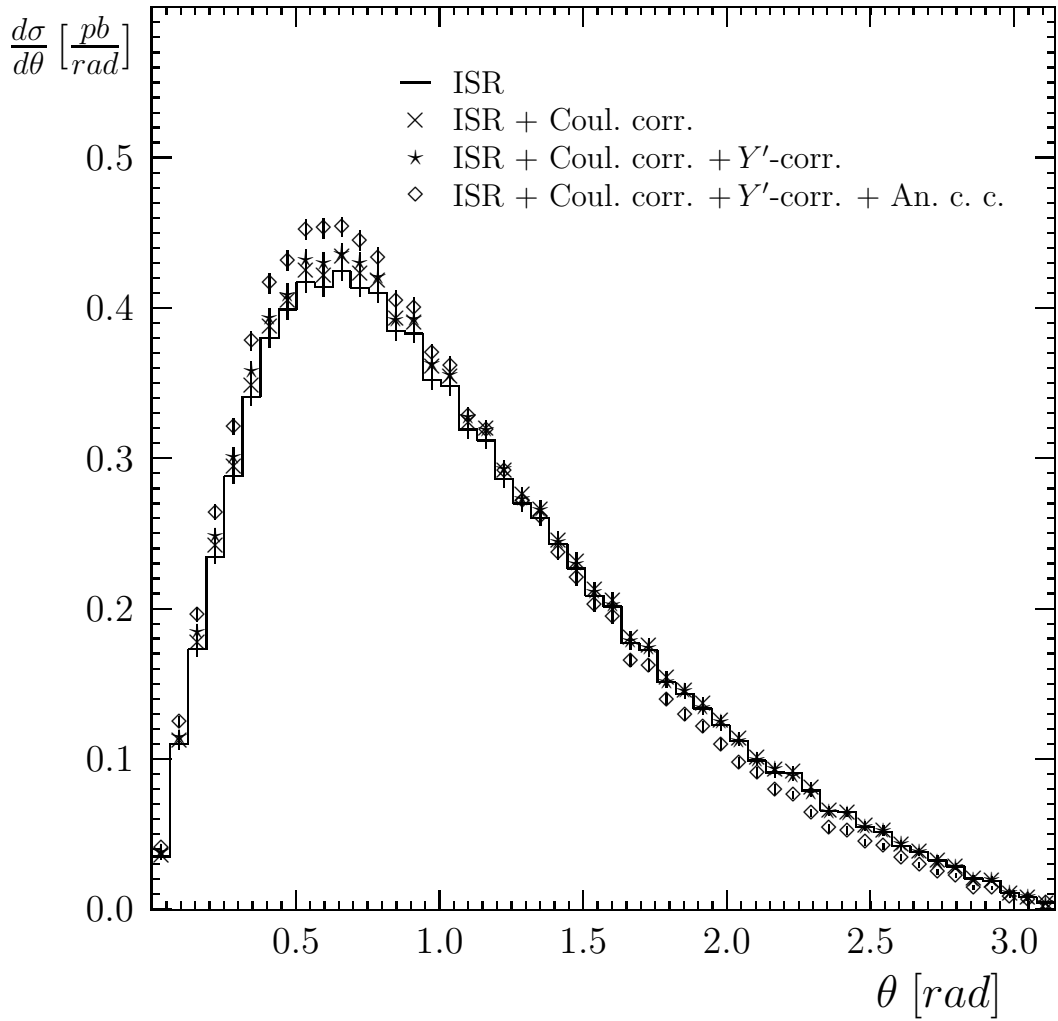

\begin{figure}[!ht]

\centering
\setlength{\unitlength}{0.1mm}
\begin{picture}(1600,1400)
\put(300,150){\begin{picture}( 1200,1200)
\put(0,0){\framebox( 1200,1200){ }}
\multiput(  190.99,0)(  190.99,0){   6}{\line(0,1){25}}
\multiput(     .00,0)(   19.10,0){  63}{\line(0,1){10}}
\multiput(  190.99,1200)(  190.99,0){   6}{\line(0,-1){25}}
\multiput(     .00,1200)(   19.10,0){  63}{\line(0,-1){10}}
\put( 191,-25){\makebox(0,0)[t]{\large $    0.5 $}}
\put( 382,-25){\makebox(0,0)[t]{\large $    1.0 $}}
\put( 573,-25){\makebox(0,0)[t]{\large $    1.5 $}}
\put( 764,-25){\makebox(0,0)[t]{\large $    2.0 $}}
\put( 955,-25){\makebox(0,0)[t]{\large $    2.5 $}}
\put(1146,-25){\makebox(0,0)[t]{\large $    3.0 $}}
\put(1046,-85){\makebox(0,0)[t]{\Large $ \theta\;[rad] $}}
\multiput(0,     .00)(0,  300.00){   5}{\line(1,0){25}}
\multiput(0,   30.00)(0,   30.00){  40}{\line(1,0){10}}
\multiput(1200,     .00)(0,  300.00){   5}{\line(-1,0){25}}
\multiput(1200,   30.00)(0,   30.00){  40}{\line(-1,0){10}}
\put(-25,   0){\makebox(0,0)[r]{\large $    0.0 $}}
\put(-25, 300){\makebox(0,0)[r]{\large $    0.2 $}}
\put(-25, 600){\makebox(0,0)[r]{\large $    0.4 $}}
\put(-25, 900){\makebox(0,0)[r]{\large $    0.6 $}}
\put(-25,1150){\makebox(0,0)[r]{\Large $ \frac{d\sigma}{d\theta}\,
                                         [\frac{pb}{rad}] $}}
\put(350,850){\begin{picture}( 300,900)
 \put(-20,250){\line(1,0){40} }
 \put(50,250){\makebox(0,0)[l]{ISR }}
 \put( 0,200){\makebox(0,0){$\times$}}                                 
 \put(50,200){\makebox(0,0)[l]{ISR $+$ Coul.~corr. }} 
 \put( 0,150){\makebox(0,0){$\star$}}                 
 \put(50,150){\makebox(0,0)[l]{ISR $+$ Coul.~corr. $+\,Y'$-corr. }}
 \put( 0,100){\makebox(0,0){$\diamond$}} 
 \put(50,100){\makebox(0,0)[l]{ISR $+$ Coul.~corr. $+\,Y'$-corr.
                               $+$ An.~c.~c. }}   
\end{picture}}                                                  
\end{picture}}
\put(300,150){\begin{picture}( 1200,1200)
\thinlines 
\newcommand{\x}[3]{\put(#1,#2){\line(1,0){#3}}}
\newcommand{\y}[3]{\put(#1,#2){\line(0,1){#3}}}
\newcommand{\z}[3]{\put(#1,#2){\line(0,-1){#3}}}
\newcommand{\e}[3]{\put(#1,#2){\line(0,1){#3}}}
\y{   0}{   0}{ 402}\x{   0}{ 402}{  24} \e{  12}{  397}{  10}
\y{  24}{ 402}{ 533}\x{  24}{ 935}{  24} \e{  36}{  927}{  16}
\z{  48}{ 935}{  86}\x{  48}{ 849}{  24} \e{  60}{  841}{  16}
\z{  72}{ 849}{ 172}\x{  72}{ 677}{  24} \e{  84}{  670}{  14}
\z{  96}{ 677}{ 126}\x{  96}{ 551}{  24} \e{ 108}{  544}{  12}
\z{ 120}{ 551}{  90}\x{ 120}{ 461}{  24} \e{ 132}{  456}{  12}
\z{ 144}{ 461}{  75}\x{ 144}{ 386}{  24} \e{ 156}{  381}{  10}
\z{ 168}{ 386}{  66}\x{ 168}{ 320}{  24} \e{ 180}{  316}{  10}
\z{ 192}{ 320}{  46}\x{ 192}{ 274}{  24} \e{ 204}{  270}{   8}
\z{ 216}{ 274}{  41}\x{ 216}{ 233}{  24} \e{ 228}{  229}{   8}
\z{ 240}{ 233}{  29}\x{ 240}{ 204}{  24} \e{ 252}{  200}{   8}
\z{ 264}{ 204}{  20}\x{ 264}{ 184}{  24} \e{ 276}{  181}{   8}
\z{ 288}{ 184}{  31}\x{ 288}{ 153}{  24} \e{ 300}{  150}{   6}
\z{ 312}{ 153}{  17}\x{ 312}{ 136}{  24} \e{ 324}{  133}{   6}
\z{ 336}{ 136}{  15}\x{ 336}{ 121}{  24} \e{ 348}{  118}{   6}
\z{ 360}{ 121}{  18}\x{ 360}{ 103}{  24} \e{ 372}{  100}{   6}
\z{ 384}{ 103}{  10}\x{ 384}{  93}{  24} \e{ 396}{   91}{   6}
\z{ 408}{  93}{   8}\x{ 408}{  85}{  24} \e{ 420}{   82}{   4}
\z{ 432}{  85}{  11}\x{ 432}{  74}{  24} \e{ 444}{   72}{   4}
\z{ 456}{  74}{   7}\x{ 456}{  67}{  24} \e{ 468}{   64}{   4}
\z{ 480}{  67}{   8}\x{ 480}{  59}{  24} \e{ 492}{   57}{   4}
\z{ 504}{  59}{   4}\x{ 504}{  55}{  24} \e{ 516}{   53}{   4}
\z{ 528}{  55}{   9}\x{ 528}{  46}{  24} \e{ 540}{   45}{   4}
\y{ 552}{  46}{   1}\x{ 552}{  47}{  24} \e{ 564}{   45}{   4}
\z{ 576}{  47}{   6}\x{ 576}{  41}{  24} \e{ 588}{   40}{   4}
\z{ 600}{  41}{   3}\x{ 600}{  38}{  24} \e{ 612}{   36}{   4}
\z{ 624}{  38}{   6}\x{ 624}{  32}{  24} \e{ 636}{   30}{   2}
\y{ 648}{  32}{   2}\x{ 648}{  34}{  24} \e{ 660}{   32}{   4}
\z{ 672}{  34}{   5}\x{ 672}{  29}{  24} \e{ 684}{   28}{   2}
\z{ 696}{  29}{   3}\x{ 696}{  26}{  24} \e{ 708}{   25}{   2}
\z{ 720}{  26}{   2}\x{ 720}{  24}{  24} \e{ 732}{   23}{   2}
\z{ 744}{  24}{   2}\x{ 744}{  22}{  24} \e{ 756}{   21}{   2}
\y{ 768}{  22}{   0}\x{ 768}{  22}{  24} \e{ 780}{   21}{   2}
\z{ 792}{  22}{   3}\x{ 792}{  19}{  24} \e{ 804}{   18}{   2}
\y{ 816}{  19}{   0}\x{ 816}{  19}{  24} \e{ 828}{   18}{   2}
\y{ 840}{  19}{   0}\x{ 840}{  19}{  24} \e{ 852}{   18}{   2}
\z{ 864}{  19}{   6}\x{ 864}{  13}{  24} \e{ 876}{   12}{   2}
\y{ 888}{  13}{   0}\x{ 888}{  13}{  24} \e{ 900}{   12}{   2}
\z{ 912}{  13}{   1}\x{ 912}{  12}{  24} \e{ 924}{   11}{   2}
\z{ 936}{  12}{   2}\x{ 936}{  10}{  24} \e{ 948}{    9}{   2}
\y{ 960}{  10}{   0}\x{ 960}{  10}{  24} \e{ 972}{    9}{   2}
\z{ 984}{  10}{   1}\x{ 984}{   9}{  24} \e{ 996}{    8}{   2}
\z{1008}{   9}{   1}\x{1008}{   8}{  24} \e{1020}{    7}{   2}
\z{1032}{   8}{   1}\x{1032}{   7}{  24} \e{1044}{    7}{   2}
\z{1056}{   7}{   1}\x{1056}{   6}{  24} \e{1068}{    6}{   2}
\z{1080}{   6}{   1}\x{1080}{   5}{  24} \e{1092}{    5}{   2}
\z{1104}{   5}{   1}\x{1104}{   4}{  24} \e{1116}{    3}{   2}
\z{1128}{   4}{   2}\x{1128}{   2}{  24} \e{1140}{    2}{   0}
\y{1152}{   2}{   0}\x{1152}{   2}{  24} \e{1164}{    2}{   0}
\z{1176}{   2}{   1}\x{1176}{   1}{  24} \e{1188}{    1}{   0}
\end{picture}} 
\put(300,150){\begin{picture}( 1200,1200)
\newcommand{\R}[2]{\put(#1,#2){\makebox(0,0){$\times$}}}
\newcommand{\e}[3]{\put(#1,#2){\line(0,1){#3}}}
\R{  12}{ 407} \e{  12}{  402}{  10} \R{  36}{ 946} \e{  36}{  938}{  16} 
\R{  60}{ 859} \e{  60}{  851}{  16} \R{  84}{ 685} \e{  84}{  678}{  14} 
\R{ 108}{ 557} \e{ 108}{  551}{  12} \R{ 132}{ 468} \e{ 132}{  462}{  12}
\R{ 156}{ 392} \e{ 156}{  386}{  10} \R{ 180}{ 325} \e{ 180}{  320}{  10}
\R{ 204}{ 277} \e{ 204}{  272}{   8} \R{ 228}{ 235} \e{ 228}{  231}{   8}
\R{ 252}{ 206} \e{ 252}{  202}{   8} \R{ 276}{ 187} \e{ 276}{  183}{   8}
\R{ 300}{ 155} \e{ 300}{  152}{   6} \R{ 324}{ 139} \e{ 324}{  135}{   6}
\R{ 348}{ 123} \e{ 348}{  120}{   6} \R{ 372}{ 104} \e{ 372}{  101}{   6}
\R{ 396}{  94} \e{ 396}{   92}{   6} \R{ 420}{  86} \e{ 420}{   83}{   4}
\R{ 444}{  76} \e{ 444}{   73}{   4} \R{ 468}{  67} \e{ 468}{   65}{   4}
\R{ 492}{  59} \e{ 492}{   57}{   4} \R{ 516}{  56} \e{ 516}{   54}{   4}
\R{ 540}{  47} \e{ 540}{   46}{   4} \R{ 564}{  47} \e{ 564}{   46}{   4}
\R{ 588}{  42} \e{ 588}{   41}{   4} \R{ 612}{  38} \e{ 612}{   36}{   4}
\R{ 636}{  32} \e{ 636}{   30}{   2} \R{ 660}{  35} \e{ 660}{   33}{   4}
\R{ 684}{  30} \e{ 684}{   28}{   2} \R{ 708}{  27} \e{ 708}{   25}{   2}
\R{ 732}{  24} \e{ 732}{   23}{   2} \R{ 756}{  23} \e{ 756}{   21}{   2}
\R{ 780}{  22} \e{ 780}{   21}{   2} \R{ 804}{  19} \e{ 804}{   18}{   2}
\R{ 828}{  19} \e{ 828}{   18}{   2} \R{ 852}{  19} \e{ 852}{   18}{   2}
\R{ 876}{  13} \e{ 876}{   12}{   2} \R{ 900}{  13} \e{ 900}{   12}{   2}
\R{ 924}{  12} \e{ 924}{   12}{   2} \R{ 948}{  10} \e{ 948}{    9}{   2}
\R{ 972}{  10} \e{ 972}{    9}{   2} \R{ 996}{   9} \e{ 996}{    8}{   2}
\R{1020}{   8} \e{1020}{    7}{   2} \R{1044}{   7} \e{1044}{    7}{   2}
\R{1068}{   7} \e{1068}{    6}{   2} \R{1092}{   5} \e{1092}{    5}{   2}
\R{1116}{   4} \e{1116}{    4}{   2} \R{1140}{   2} \e{1140}{    2}{   0}
\R{1164}{   2} \e{1164}{    2}{   0} \R{1188}{   1} \e{1188}{    1}{   0}
\end{picture}} 

\put(300,150){\begin{picture}( 1200,1200)
\newcommand{\R}[2]{\put(#1,#2){\makebox(0,0){$\star$}}}
\newcommand{\e}[3]{\put(#1,#2){\line(0,1){#3}}}
\R{  12}{ 427} \e{  12}{  422}{  12} \R{  36}{ 980} \e{  36}{  971}{  16} 
\R{  60}{ 874} \e{  60}{  866}{  16} \R{  84}{ 700} \e{  84}{  693}{  14}
\R{ 108}{ 558} \e{ 108}{  552}{  12} \R{ 132}{ 469} \e{ 132}{  463}{  12}
\R{ 156}{ 388} \e{ 156}{  383}{  10} \R{ 180}{ 326} \e{ 180}{  321}{  10}
\R{ 204}{ 273} \e{ 204}{  269}{   8} \R{ 228}{ 236} \e{ 228}{  232}{   8}
\R{ 252}{ 209} \e{ 252}{  205}{   8} \R{ 276}{ 183} \e{ 276}{  180}{   8}
\R{ 300}{ 151} \e{ 300}{  148}{   6} \R{ 324}{ 136} \e{ 324}{  132}{   6}
\R{ 348}{ 122} \e{ 348}{  119}{   6} \R{ 372}{ 102} \e{ 372}{   99}{   6}
\R{ 396}{  92} \e{ 396}{   89}{   6} \R{ 420}{  85} \e{ 420}{   83}{   4}
\R{ 444}{  76} \e{ 444}{   73}{   4} \R{ 468}{  67} \e{ 468}{   65}{   4}
\R{ 492}{  57} \e{ 492}{   55}{   4} \R{ 516}{  55} \e{ 516}{   53}{   4}
\R{ 540}{  47} \e{ 540}{   45}{   4} \R{ 564}{  46} \e{ 564}{   44}{   4}
\R{ 588}{  43} \e{ 588}{   41}{   4} \R{ 612}{  38} \e{ 612}{   37}{   4}
\R{ 636}{  33} \e{ 636}{   31}{   4} \R{ 660}{  33} \e{ 660}{   32}{   4}
\R{ 684}{  29} \e{ 684}{   28}{   2} \R{ 708}{  26} \e{ 708}{   25}{   2}
\R{ 732}{  25} \e{ 732}{   23}{   2} \R{ 756}{  22} \e{ 756}{   21}{   2}
\R{ 780}{  22} \e{ 780}{   21}{   2} \R{ 804}{  19} \e{ 804}{   18}{   2}
\R{ 828}{  21} \e{ 828}{   19}{   2} \R{ 852}{  18} \e{ 852}{   17}{   2}
\R{ 876}{  13} \e{ 876}{   12}{   2} \R{ 900}{  13} \e{ 900}{   12}{   2}
\R{ 924}{  11} \e{ 924}{   10}{   2} \R{ 948}{  10} \e{ 948}{   10}{   2}
\R{ 972}{  10} \e{ 972}{    9}{   2} \R{ 996}{   9} \e{ 996}{    8}{   2}
\R{1020}{   7} \e{1020}{    7}{   2} \R{1044}{   7} \e{1044}{    6}{   2}
\R{1068}{   7} \e{1068}{    6}{   2} \R{1092}{   5} \e{1092}{    5}{   2}
\R{1116}{   4} \e{1116}{    4}{   2} \R{1140}{   3} \e{1140}{    2}{   0}
\R{1164}{   2} \e{1164}{    2}{   0} \R{1188}{   1} \e{1188}{    0}{   0}
\end{picture}} 
\put(300,150){\begin{picture}( 1200,1200)
\newcommand{\R}[2]{\put(#1,#2){\makebox(0,0){$\diamond$}}}
\newcommand{\e}[3]{\put(#1,#2){\line(0,1){#3}}}
\R{  12}{ 442} \e{  12}{  435}{  12} \R{  36}{ 992} \e{  36}{  982}{  18}
\R{  60}{ 889} \e{  60}{  881}{  18} \R{  84}{ 715} \e{  84}{  707}{  16}
\R{ 108}{ 561} \e{ 108}{  554}{  14} \R{ 132}{ 478} \e{ 132}{  471}{  12}
\R{ 156}{ 397} \e{ 156}{  391}{  12} \R{ 180}{ 337} \e{ 180}{  331}{  10}
\R{ 204}{ 286} \e{ 204}{  281}{  10} \R{ 228}{ 249} \e{ 228}{  245}{  10}
\R{ 252}{ 232} \e{ 252}{  227}{   8} \R{ 276}{ 207} \e{ 276}{  203}{   8}
\R{ 300}{ 180} \e{ 300}{  176}{   8} \R{ 324}{ 166} \e{ 324}{  162}{   8}
\R{ 348}{ 155} \e{ 348}{  151}{   8} \R{ 372}{ 144} \e{ 372}{  141}{   8}
\R{ 396}{ 137} \e{ 396}{  134}{   6} \R{ 420}{ 136} \e{ 420}{  132}{   6}
\R{ 444}{ 127} \e{ 444}{  124}{   6} \R{ 468}{ 123} \e{ 468}{  119}{   6}
\R{ 492}{ 117} \e{ 492}{  114}{   6} \R{ 516}{ 114} \e{ 516}{  111}{   6}
\R{ 540}{ 112} \e{ 540}{  109}{   6} \R{ 564}{ 110} \e{ 564}{  107}{   6}
\R{ 588}{ 110} \e{ 588}{  107}{   6} \R{ 612}{ 106} \e{ 612}{  103}{   6}
\R{ 636}{ 100} \e{ 636}{   98}{   6} \R{ 660}{  96} \e{ 660}{   93}{   6}
\R{ 684}{  81} \e{ 684}{   79}{   6} \R{ 708}{  85} \e{ 708}{   82}{   6}
\R{ 732}{  77} \e{ 732}{   75}{   6} \R{ 756}{  68} \e{ 756}{   65}{   4}
\R{ 780}{  63} \e{ 780}{   61}{   4} \R{ 804}{  62} \e{ 804}{   60}{   4}
\R{ 828}{  56} \e{ 828}{   54}{   4} \R{ 852}{  48} \e{ 852}{   46}{   4}
\R{ 876}{  38} \e{ 876}{   37}{   4} \R{ 900}{  36} \e{ 900}{   34}{   4}
\R{ 924}{  31} \e{ 924}{   29}{   4} \R{ 948}{  27} \e{ 948}{   25}{   4}
\R{ 972}{  21} \e{ 972}{   20}{   2} \R{ 996}{  16} \e{ 996}{   14}{   2}
\R{1020}{  14} \e{1020}{   13}{   2} \R{1044}{   9} \e{1044}{    8}{   2}
\R{1068}{  10} \e{1068}{    9}{   2} \R{1092}{   6} \e{1092}{    5}{   2}
\R{1116}{   4} \e{1116}{    4}{   2} \R{1140}{   3} \e{1140}{    3}{   2}
\R{1164}{   2} \e{1164}{    2}{   0} \R{1188}{   1} \e{1188}{    0}{   0}
\end{picture}} 

\end{picture} 

\caption{\small\sf $W^-$ angular distributions for $E_{CM}=500\,GeV$. 
See the text for more details.
}
\label{fig:e500}
\end{figure}
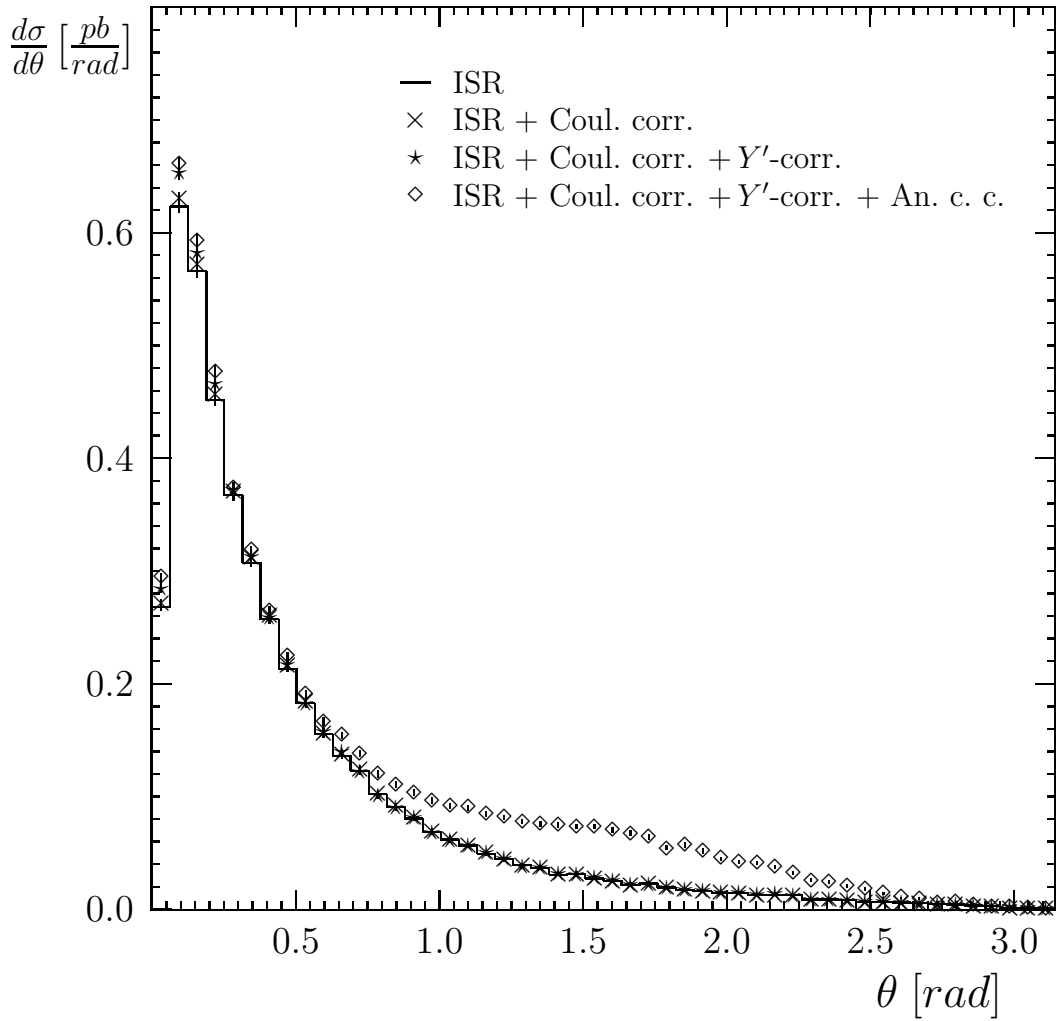

What we see is that the effect of the full YFS form factor is at the
level of 
$0.14\%$, $0.39\%$ and $0.52\%$, 
respectively beyond the usual Coulomb effect at the  
LEP2 energies $175\,GeV$, $190\,GeV$ and $205\,GeV$ and is at the level of 
$0.81\%$ beyond the usual Coulomb effect
at NLC energies for the case of the SM couplings, giving a total effect
beyond initial state radiation of 
$3.00\%$, $2.59\%$, $2.44\%$ and $2.06\%$
for the CMS energies $175\,GeV$, $190\,GeV$, $205\,GeV$ and $500\,GeV$, 
respectively. For the case of the anomalous couplings of 
$\delta\kappa=\delta\lambda=0.1$,
the corresponding results are 
$0.22\%$, $0.47\%$, $0.62\%$ and $0.25\%$, 
respectively beyond the usual Coulomb effect for total corrections of 
$3.06\%$, $2.67\%$, $2.55\%$ and $1.50\%$, 
respectively for the same CMS energies.
Thus, we see that at LEP2 and NLC energies, the full form factor
does indeed modulate the effect of the anomalous couplings; 
this enhances its importance at both LEP2 and NLC energies. Since the 
targeted accuracy of the theoretical precision for the 
LEP2 $WW$-pair production cross section is  $0.5\%$ \cite{WWWG},
evidently, the full form factor effect is very important for LEP2
physics scenarios.

Indeed, we have looked into the manifestation of these effects
in the $W^{\pm}$ production angular distributions. We show in 
Figs.~\ref{fig:e190},~\ref{fig:e500}
four respective differential distributions for the total
cross sections in Table~\ref{tab:xsec} for the LEP2 energy of 
$\sqrt s = 190\,GeV$ and for the NLC energy of $\sqrt s = 500\,GeV$, where we
feature the $W^-$ production angle distribution in the CM system for
all three SM coupling scenarios in Table~\ref{tab:xsec} and with the final
``ISR+Coul. corr.+Y$'$-corr.'' scenario for the anomalous coupling case.
We see in these figures that the full formfactor effect modulates
the distributions most strongly near their peaks, near the forward 
direction for the $W^-$ when the incoming electron direction is used as the 
reference direction. As the anomalous couplings modulate these distributions
over a large range of the respective production angles, we see 
that the smaller (larger) values of the full formfactor effects in 
Table~\ref{tab:xsec}
for the anomalous cases relative to the SM cases
for NLC (LEP2) energies is consistent
with the shapes of the distributions in Figs.~\ref{fig:e190},~\ref{fig:e500}. 
Also evident in the figures is the more pronounced anomalous coupling effect 
at NLC energies, as expected.
\par
We end this section by noting that we have also implemented the 
full YFS form factor effect and 
the anomalous couplings as well in an unpublished version of the program
KORALW~\cite{KORALW} of three of us (M.S., S.J. and W.P.) and we have
checked that the results from YFSWW2 and from this new version of KORALW
~\cite{KORALWN} are in agreement within the statistical errors
of the simulations. This is an important cross-check on the results
in this paper it will be presented in detail elsewhere~\cite{KORALWN}.
\par
\section{ Conclusions}

In this paper we have developed the YFS theory of charged, spin 1 bosons
in the presence of possible non-zero widths. We have applied our theory
to the LEP2/NLC process $e^+e^-\rightarrow W^+W^-+n(\gamma)$, allowing for
the W-pair to decay to 4 fermions. The result is the description, via the
Monte Carlo event generator YFSWW2, of the respective $n(\gamma)$
effects on an event-by-event basis, in which the infrared singularities
are cancelled to all orders in $\alpha$.

Specifically, in our realization of the YFS theory for charged spin~1 bosons, 
we have maintained the electromagnetic
gauge invariance of the $SU_{2L}\times U_1$ theory following the
works in Refs.~\cite{B-Z,argy}. In addition, we have also avoided
any doubling counting of the so-called Coulomb effect by removing
it analytically from the YFS virtual infrared function $B$. This
resulted in the definition of a new YFS virtual infrared function $B'$.
We have illustrated our calculations with explicit Monte Carlo data
at the LEP2 CMS energies $\sqrt s = 175\,GeV,\,190\,GeV,$ and $205\,GeV$
and at the NLC energy $\sqrt s = 500\,GeV$, for both Standard Model and
anomalous $VWW$ vertices, $V= \gamma, Z$. We find in all cases that
the effect of the radiation by the $W^{\pm}$ themselves is important,
both in the production angular distributions and in the over-all 
normalization.
In our YFS Monte Carlo realization of this effect, we have worked to
the leading $\bar\beta_0$ level in the current analysis. The 
higher order hard photon residuals $\bar\beta_n$, $n\ge 1$
will be considered elsewhere~\cite{elsewh} in this connection.

In summary, our Monte Carlo event generator YFSWW2 now calculates,
on an event-by-event basis, the multiple photon effects in the
process $e^+e^-\rightarrow W^+W^-+n(\gamma) \rightarrow 4 fermions +n(\gamma)$
and includes, for the first time ever, the effects of the radiation
by the $W^\pm$ themselves in the respective YFS exponentiated soft
photons, without doubling counting the so-called Coulomb effect
and without spoiling the electromagnetic gauge invariance of the
$SU_{2L}\times U_1$ theory. This program is available from the authors
at the WWW URL http://enigma.phys.utk.edu/pub/YFSWW/
and we look forward with excitement to its application to imminent LEP2
data.

\vspace{7mm}
\noindent 
{\large\bf  Acknowledgements}

Two of us (S. J. and B.F.L. W.) acknowledge the
kind hospitality of Prof. G. Veneziano and the CERN Theory 
Division while this work was completed. We would like to acknowledge
the helpful discussion of E. Richter-Was in the calculation of YFS
form factors.

\appendix

\section{YFS infrared functions}

In this section we present some analytical representations of
the YFS infrared (IR) functions corresponding to emission of virtual
and real photons for the $W^+W^-$-pair production in the $e^+e^-$
annihilation. An important feature of these representations is
that they are stable and fast in numerical evaluation. Thus, they 
are particularly suited for Monte Carlo implementations.     

\subsection{Virtual photon IR function for $s$-channel}

The $s$-channel virtual photon YFS IR function $\Re B(s)$ for any
two charged particles with four-momenta $p_1,p_2$ and
masses $m_1,m_2$ reads
\begin{align}
2\alpha\Re B(s,m_1,m_2) = \frac{\alpha}{\pi}\,\Bigg\{
& \left(\frac{1}{\rho}\ln\frac{\mu(1+\rho)}{m_1m_2} -1 \right)
  \ln\frac{m_{\gamma}^2}{m_1m_2} 
  + \frac{\mu\rho}{s}\ln\frac{\mu(1+\rho)}{m_1m_2}
  + \frac{m_1^2-m_2^2}{2s}\ln\frac{m_1}{m_2} 
\notag\\
+ & \frac{1}{\rho} \Bigg[\,\pi^2 
  - \frac{1}{2}\ln\frac{\mu(1+\rho)}{m_1^2} \ln\frac{\mu(1+\rho)}{m_2^2}
  - \frac{1}{2}\ln^2\frac{m_1^2+\mu(1+\rho)}{m_2^2+\mu(1+\rho)}
\notag\\  
& \;- Li_2\left(\frac{2\mu\rho}{m_1^2+\mu(1+\rho)}\right) 
    - Li_2\left(\frac{2\mu\rho}{m_2^2+\mu(1+\rho)}\right)
                   \Bigg] - 1                            \Bigg\},
\label{ReBs}
\end{align}
where
\begin{align}
& \mu=p_1p_2, \:\:\: s=2\mu + m_1^2 + m_2^2,
\notag \\
& \rho=\sqrt{1 - \left(\frac{m_1m_2}{\mu}\right)^2},
\label{mu-rho}
\end{align}
and $m_{\gamma}$ is a fictitious photon mass used to regularize
the IR singularity. 

\subsection{Virtual photon IR function for $t$ and $u$ -channels}

The $t$-channel virtual photon YFS IR function $\Re B(t)$ for 
two charged particles with four-momenta $p_1,p_2$ and
masses $m,M$, where $m\ll M,|t|$, reads
\begin{align}
2\alpha\Re B(t,m,M) = \frac{\alpha}{\pi}\,\Bigg\{
& \left(\ln\frac{|t|}{mM} + \ln\zeta - 1\right)\ln\frac{m_{\gamma}^2}{m^2}
  + \frac{\zeta}{2}\left(\ln\frac{|t|}{mM} + \ln\zeta\right)
\notag\\
& -\frac{1}{2}\ln\frac{|t|}{m^2}\ln\frac{|t|}{M^2}
  -\ln\frac{M}{m}\left(\ln\frac{|t|}{mM} + \ln\zeta + \frac{\zeta-3}{2}\right)
\notag\\
& -\ln\zeta\left(\ln\frac{|t|}{mM} + \frac{1}{2}\ln\zeta\right)
  +Li_2\left(\frac{1}{\zeta}\right) - 1                   \Bigg\},
\label{ReBt}
\end{align}
where
\begin{equation}
\zeta = 1 + \frac{M^2}{|t|}\,, \:\:\: t = (p_1-p_2)^2.
\label{zeta-t}
\end{equation}
The $u$-channel IR function $\Re B(u)$ can be obtained simply by
replacing $t\rightarrow u$ in the above formula.

\subsection{Real photon IR function}
The YFS IR function $\tilde{B}$ corresponding to the emission of real photons
with energy $E_{\gamma}\le K_{max}$ in a process involving any two charged 
particles with four-momenta $p_1,p_2$ and masses $m_1,m_2$ can be expressed as 
\begin{align}
2\alpha\tilde{B}(p_1,p_2;K_{max}) = \frac{\alpha}{\pi}\,\Bigg\{
& \left(\frac{1}{\rho}\ln\frac{\mu(1+\rho)}{m_1m_2} -1 \right)  
  \ln\frac{4K_{max}^2}{m_{\gamma}^2 }
  + \frac{1}{2\beta_1}\ln\frac{1+\beta_1}{1-\beta_1} 
  + \frac{1}{2\beta_2}\ln\frac{1+\beta_2}{1-\beta_2}
\notag\\
& + \mu A_4(p_1,p_2)                                                 \Bigg\},
\label{B-tilde}
\end{align}
where $\beta_i=\sqrt{1-m_i^2/E_i^2}$, and $\mu$ and $\rho$ are defined in
Eq.~(\ref{mu-rho}). The most complicated part of the above expression
is the function $A_4(p_1,p_2)$. It can be expressed as a combination
of some number of logarithms and dilogarithms\footnote{By using some
identities we managed to reduce the number of dilogarithms to 8 only!}
(see also Ref.~\cite{sj-zw} for a similar calculation)
\begin{align}
A_4(p_1,p_2) = \frac{1}{\sqrt{(Q^2+\omega^2)(Q^2+\delta^2)}} \Big\{
& \ln\frac{\sqrt{\Delta^2+Q^2}-\Delta}{\sqrt{\Delta^2+Q^2}+\Delta}
  \left[X_{23}^{14}(\eta_1) - X_{23}^{14}(\eta_0)\right]
\notag\\
& + Y(\eta_1) - Y(\eta_0)                                    \Big\},
\label{A4} 
\end{align}
where
\begin{align}
& X_{kl}^{ij}(\eta) = 
  \ln\left|\frac{(\eta-y_i)(\eta-y_j)}{(\eta-y_k)(\eta-y_l)}\right|,
\notag\\
& Y(\eta) = Z_{14}(\eta) + Z_{21}(\eta) + Z_{32}(\eta) - Z_{34}(\eta)
          + \frac{1}{2}X_{34}^{12}(\eta)X_{14}^{23}(\eta),
\label{XYZ} \\
& Z_{ij}(\eta) = 2\Re Li_2\left(\frac{y_j-y_i}{\eta-y_i}\right)
               + \frac{1}{2}\ln^2\left|\frac{\eta-y_i}{\eta-y_j}\right|,
\notag
\end{align}
and
\begin{align}
&\eta_0=\sqrt{E_2^2-m_2^2}, \:\:\:\:\:\: 
\eta_1=\sqrt{E_1^2-m_1^2} + \sqrt{\Delta^2+Q^2},
\notag\\
& y_{1,2} = \frac{1}{2}\left[\sqrt{\Delta^2+Q^2}-\Omega 
          + \frac{\omega\delta\pm\sqrt{(Q^2+\omega^2)(Q^2+\delta^2)}}
                 {\sqrt{\Delta^2+Q^2}+\Delta}
                       \right],
\label{eta-y}\\
& y_{3,4} = \frac{1}{2}\left[\sqrt{\Delta^2+Q^2}+\Omega 
          + \frac{\omega\delta\pm\sqrt{(Q^2+\omega^2)(Q^2+\delta^2)}}
                 {\sqrt{\Delta^2+Q^2}-\Delta}
                       \right],
\notag
\end{align}
where we used the following notation
\begin{align}
& \Delta = E_1 - E_2, \:\:\:\:\: \Omega = E_1 + E_2,
\notag\\
& \delta = m_1 - m_2, \:\:\:\:\: \omega = m_1 + m_2,
\label{Bt-not}\\
& Q^2 = -(p_1-p_2)^2.
\notag
\end{align}
The only approximation used in deriving the above formulae is  
$m_{\gamma}\ll K_{max}$.

As one can easily check the dependence of the above functions
on the IR regulator $m_{\gamma}$ cancels out in the sum 
$2\alpha\Re B + 2\alpha\tilde{B}$ which is used to construct the
YFS formfactor.


 
\end{document}